\documentclass[conference]{IEEEtran}
\usepackage[left=1.57cm,right=1.57cm,top=2.5cm,bottom=3cm,headsep=1cm,footskip=1.5cm]{geometry}
\IEEEoverridecommandlockouts
\usepackage{cite}
\usepackage{amsmath,amssymb,amsfonts}
\renewcommand\IEEEkeywordsname{Keywords}
\usepackage{algorithmic}
\usepackage{graphicx}
\usepackage{textcomp}
\usepackage{xcolor}
\def\BibTeX{{\rm B\kern-.05em{\sc i\kern-.025em b}\kern-.08em
    T\kern-.1667em\lower.7ex\hbox{E}\kern-.125emX}}

\usepackage{url}
\usepackage{hyperref}
\hypersetup{
    colorlinks=true,
    linkcolor=blue,
    citecolor=blue,
    filecolor=magenta,      
    urlcolor=cyan,
}
\usepackage{blkarray}
\usepackage{mathtools}
\usepackage{algorithm2e}
\RestyleAlgo{ruled}
\usepackage{subcaption}
\usepackage[font=footnotesize]{caption}
\DeclareMathOperator{\adj}{adj}
\usepackage{stfloats}

\usepackage{fancyhdr}

\fancypagestyle{firstpage}{
\fancyhf{}
\fancyhead[L]{Journal of Robotics and Control (JRC)\\
Volume 4, Issue 1, January 2023\\
ISSN: 2715-5072, DOI: 10.18196/jrc.v4i1.16164}
\fancyhead[R]{\thepage}          
\fancyfoot[L]{\includegraphics{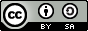}}
\fancyfoot[C]{Journal Web site: \url{http://journal.umy.ac.id/index.php/jrc}}
\fancyfoot[R]{Journal Email: jrc@umy.ac.id}

}

\fancypagestyle{followingpage}{
\fancyhf{}
\fancyhead[L]{Journal of Robotics and Control (JRC)}
\fancyhead[C]{ISSN: 2715-5072}
\fancyhead[R]{\thepage}
\fancyfoot[L]{Author Name, Article Title}

}

\AtBeginDocument{\thispagestyle{firstpage}}
\begin{document}

\title{Non-Linear Estimation using the Weighted Average Consensus-Based Unscented Filtering for Various Vehicles Dynamics towards Autonomous Sensorless Design
}

\author{Bambang L. Widjiantoro \textsuperscript{1}, Moh Kamalul Wafi \textsuperscript{2*}, Katherin Indriawati \textsuperscript{3}\\
\textsuperscript{1,2,3} Department of Engineering Physics, Institut Teknologi Sepuluh Nopember, Surabaya, Indonesia \\
Email: \textsuperscript{1} blelono@ep.its.ac.id, \textsuperscript{2} kamalul.wafi@its.ac.id, \textsuperscript{3} katherin@ep.its.ac.id\\
*Corresponding Author
}

\maketitle

\begin{abstract}
The concerns to autonomous vehicles have been becoming more intriguing in coping with the more environmentally dynamics non-linear systems under some constraints and disturbances. These vehicles connect not only to the self-instruments yet to the neighborhoods components, making the diverse interconnected communications which should be handled locally to ease the computation and to fasten the decision. To deal with those interconnected networks, the distributed estimation to reach the untouched states, pursuing sensorless design, is approached, initiated by the construction of the modified pseudo measurement which, due to approximation, led to the weighted average consensus calculation within unscented filtering along with the bounded estimation errors. Moreover, the tested vehicles are also associated to certain robust control scenarios subject to noise and disturbance with some stability analysis to ensure the usage of the proposed estimation algorithm. The numerical instances are presented along with the performances of the control and estimation method. The results affirms the effectiveness of the method with limited error deviation compared to the other centralized and distributed filtering. Beyond these, the further research would be the directed sensorless design and fault-tolerant learning control subject to faults to negate the failures.
\end{abstract}

\section{Introduction}
The ideas of autonomous vehicles being operated in certain environment have been discussed before the 20th century \cite{R1}, from underwater, land, and air, however the future challenges with advanced manufacturing are constantly open to tackle \cite{R2} in addition to the current surveys \cite{R3}. The breakthrough in communication technologies along with the more dynamics entities causes this subject more reality with the guarantees from the stability studies, such as using neural-network \cite{R4} and under some regular switching methods \cite{R5}. Moreover, the stability of the discretized autonomous system was also studied \cite{R6} in a connected networked system \cite{R7} subject to delays while considering the stability controls \cite{R8,R9}. To give more comprehensive results, the examined autonomous vehicles in this paper comprise various dynamical systems; the adaptive cruise control \cite{R10,R11}, the active suspension system \cite{R12,R13}, the electric aircraft \cite{R14,R15}, and the DC motor drive system \cite{R16,R17}. To deal with the more interconnected systems, the control design is required from the basic feedback control \cite{R18} to intelligent-based control \cite{R19,R20} in real-time \cite{R21} with constraints \cite{R22}.

The stability analysis and control designs, as the discussions in the paper, then leads to the modest-cost implementation using sensorless design in lieu of the hardware-sensor from the basic estimation methods. These methods vary in applications, comprising in standard filtering \cite{R23}, the adaptive estimation method \cite{R24}, and the most current distributed estimation \cite{R25}, while this paper focuses on the unscented Kalman filtering (UKF) as the foundation. This UKF estimation, said to answer the lack of EKF estimation in terms of the Gaussian Random variable (GRV), has been widely applied in non-linear systems \cite{R26} and the adaptive dynamical systems \cite{R27} with considering the stochastic uncertainties \cite{R28}. However, this sensorless design is supposed to be distributed, therefore the modified UKF estimation, which is the one applied in this paper, is required to construct as written in \cite{R30} with sensor networks and it would be propagated into the mentioned vehicles. The pseudo measurement matrix presented in \cite{R31} was the basis of this development mentioning the Markov non-linear system while this was the linearized approximation \cite{R32} and the ultimate consensus-based algorithms in multi-vehicle implementing cooperative control were studied in \cite{R33} and also for the sake of distributed filtering \cite{R34}.

From those, this research focuses to study the effectiveness of the proposed distributed estimation into various vehicles as the basis of sensorless networks which could be developed into the simpler relaxed computation \cite{R29} to ease the decision. Beyond that, the simplified robust sensorless algorithms in electric vehicles and drives are our upcoming concerns of research in addition to fault-tolerant distributed learning control taking into account these research from \cite{R35,R36,R37,R38,R39} and \cite{R40,R41,R42,R43,R44,R45,R46,R47}. 

\newpage
\section{Mathematical Dynamics}\label{Sec2}
This section focuses on building the dynamics vehicle-related systems being used the examine the effectiveness of the proposed scenarios. There are five various vehicle plants, from the car cruise control, the quarter bus-suspension with disturbance, the longitudinal-pitch of the aeroplane, to the speed and position of DC motor. Beyond that, the control mechanism along with the stability discussion under some limited assumptions are written in the following sections before the algorithms are then proposed, leading to the sensorless designs.

\subsection{The Car Cruise Control Model}
In many recent contemporary vehicles, the development of any advanced self-acting controls is demanding to guarantee the safety alongside the comfort, no exception to cruise control. This cruise control is designed to maintain a stable desired speed regardless arbitrary disturbances, including the alterations of winds and roads.
\begin{figure}[h!]
    \centering
    \includegraphics[width=.4\textwidth]{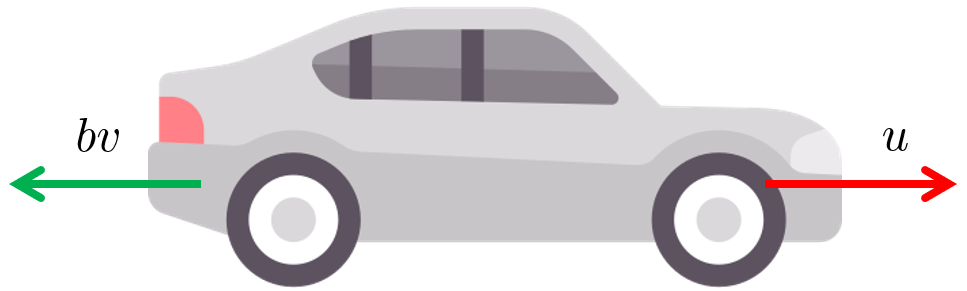}
    \caption{Free-body dynamic of the car}
    \label{F1}
\end{figure}
Moreover, it is then achieved by comparing the measured and the desired speed, which regulating the throttle based on the designed control scenario. The dynamic of the vehicle ($m$) is depicted in Fig.(\ref{F1}) with the force ($u$) being produced at the road surface, by assuming the perfect control to this force and neglecting the arbitrary forces acting on producing the force. By contrast on the mode's motion, the resistive external forces ($bv$) are implied to be linearly changed with respect to the velocity ($v$). Eq.(\ref{Eq1}) is the dynamic of the Newton's second law along with the measured system ($y$),  
\begin{gather}
    m\dot{v} + bv = u, \qquad \textrm{and} \qquad y = v \label{Eq1}
\end{gather}
and the basic state-space representative constitute as follows,
\begin{align}\begin{aligned}
    \dot{\textbf{x}} = \left[\dot{v}\right] &= \left[\frac{-b}{m}\right]v + \left[\frac{1}{m}\right]u, \\
    y &= 1\cdot v \end{aligned} \label{Eq2}
\end{align}
which is from Eq.(\ref{Eq2}), the transfer function, $\Phi_n(s), \forall n = 1, \dots$ showing the order of the examined systems, results in Eq.(\ref{Eq3}),
\begin{align}
    \Phi_1(s) = \frac{V(s)}{U(s)} = \frac{1}{ms + b} \frac{m}{Ns} \label{Eq3}
\end{align}

\subsection{A Quarter Bus-Suspension Design}
The attractive advanced suspension designs are becoming more intriguing, being linked to autonomous design. This active suspension scenario, with actuator enabling to produce the control force ($u$) acting on the body motion control, is derived from a-quarter simplified bus-mode design. The variables are explained as follows; $M_1$ and $M_2$ denote the a-quarter body and suspension mass, the constants of springs ($k_n$) and dampers ($b_n$), $\forall n = 1,2$, of suspension and wheel in turn along with non-linear disturbance ($\gamma$). From Newton's law the equations of motions could be written as Eq.(\ref{Eq4}),
\begin{figure}
    \centering
    \includegraphics[width=.43\textwidth]{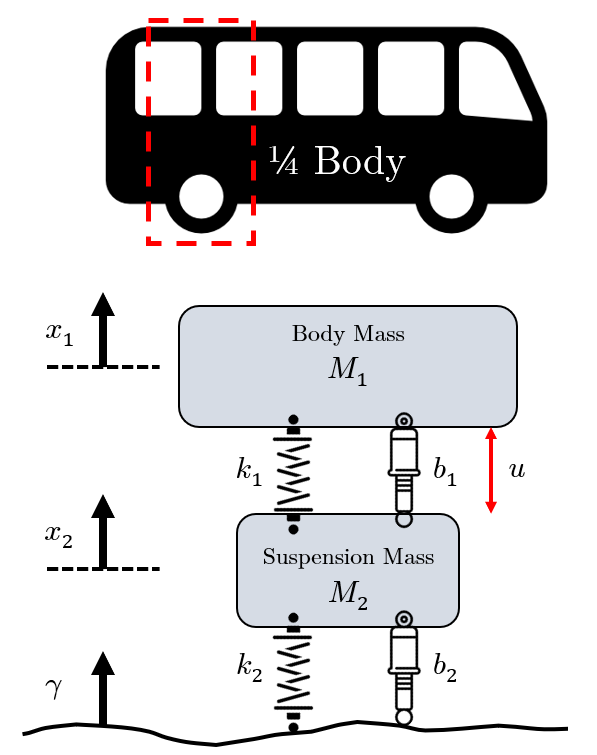}
    \caption{A quarter bus suspension}
    \label{F2}
\end{figure}
\begin{align}\begin{gathered}
    M_1\ddot{x}_1 = -\psi_{b_1} - \psi_{k_1} + u \\
    M_2\ddot{x}_2 = \psi_{b_1} + \psi_{k_1} + \psi_{b_2} + \psi_{k_2} - u \end{gathered} \label{Eq4}
\end{align}
with $\psi_\bullet$ defines the associated dynamics of ($\bullet$)-term, such that
\begin{align*}\begin{aligned}
    \psi_{b_1} &= b_1\left(\dot{x}_1 - \dot{x}_2\right) \quad&\quad \psi_{b_2} &= b_2\left(\dot{\gamma} - \dot{x}_2\right)\\
    \psi_{k_1} &= k_1(x_1 - x_2) \quad&\quad \psi_{k_2} &= k_2(\gamma - x_2)\end{aligned}
\end{align*}
and the Laplacian functions, with zero initial conditions along with the input ($u,\gamma$) and the output ($x_1-x_2$) are drawn as,
\begin{gather*}
    M_1s^2 X_1(s) + \Psi_{b_1}(s) + \Psi_{k_1}(s) = U(s)\\
    M_2s^2 X_2(s) - \Psi_{b_1}(s) - \Psi_{k_1}(s) - \Psi_{b_2}(s) - \Psi_{k_2}(s) = -U(s)
\end{gather*}
which could be concluded into standard algebraic equation,
\begin{align}
    \textbf{F}x = \textbf{g} \label{Eq5}
\end{align}
where the terms of $\textbf{F},x$, and $\textbf{g}$ in Eq.(\ref{Eq5}) are made of,
\begin{align*}
    \textbf{F} &= \begin{bmatrix}
    M_1s^2 + b_1s + k_1 & -(b_1s + k_1)\\
    -(b_1s + k_1) & M_2s^2 + (b_1 + b_2)s + (k_1 + k_2)
    \end{bmatrix}\\
    x &= \begin{bmatrix}
    X_1(s) \\
    X_2(s)
    \end{bmatrix}, \qquad\textrm{and}\qquad\, \textbf{g} = \begin{bmatrix}
    U(s) \\
    (b_2s + k_2)\Gamma(s) - U(s)
    \end{bmatrix}
\end{align*}
and the value of $x$ is obtained from the inverse with slight modification of matrix ($g$) such that it only appears $U(s)$ and $\Gamma(s)$ as in Eq.(\ref{Eq6}) with the detail parameters of $\Delta_n$, 
\begin{align}
    x = \frac{1}{\det(\textbf{F})} \begin{bmatrix}
    \Delta_1(s) & \Delta_2(s)\\
    \Delta_3(s) & \Delta_4(s)
    \end{bmatrix}\begin{bmatrix}
    U(s)\\
    \Gamma(s)
    \end{bmatrix} \label{Eq6}
\end{align}
where, after being altered, the $\adj(\textbf{F})\times \textbf{g}$ is then turned into $\Delta_n, \forall n=1\to 4$ as described below,
\begin{align*}
    \Delta_1(s) &= M_2s^2 + b_2s + k_2 & \Delta_2(s) &= z(s)\\
    \Delta_4(s) &= M_1b_2s^3 + M_1k_2s^2 + z(s) & \Delta_3(s) &= -M_1s^2
\end{align*}
with the modification of initial $\textbf{g}$ into the remaining $U(s), \Gamma(s)$
\begin{align*}
    \begin{bmatrix}
    U(s)\\
    (b_2s+k_2)\Gamma(s)- U(s)
    \end{bmatrix} \quad \longrightarrow\quad \begin{bmatrix}
    U(s)\\
    \Gamma(s)
    \end{bmatrix}
\end{align*}
where $z(s)$ in $\Delta_2(s)$ term is $b_1b_2s^2 + (b_1k_2 + b_2k_1)s + k_1k_2$. To construct the transfer functions $\Phi_2(s)$, it is required to set the sequence of the inputs. For $\Phi_{2a}$, the control input ($u$) is taken and the disturbance ($\gamma$) is assumed zero while for $\Phi_{2b}$ is the reciprocal as in Eq.(\ref{Eq7}) in turn, 
\begin{align}\begin{aligned}
    \Phi_{2a}(s) &= \frac{X_1(s) - X_2(s)}{U(s)} \\&= \frac{(M_1 + M_2)s^2 + b_2s + k_2}{\det(\textbf{F})}&\rightarrow\gamma = 0\\
    \Phi_{2b}(s) &= \frac{X_1(s) - X_2(s)}{\Gamma(s)} \\&= \frac{-M_1b_2s^3 - M_1k_2s^2}{\det(\textbf{F})}&\rightarrow u = 0\end{aligned}\label{Eq7}
\end{align}
Beyond that, the state-space representative is shown in Eq.(\ref{Eq8}) with the respected extended matrices in Eq.(\ref{Eq9}). Furthermore, the state variables includes ($x_1, y_1$) and their derivative with $y_1 = x_1 - x_2$ while the output $y = y_1$, therefore
\begin{align}\begin{aligned}
    \dot{\textbf{x}} &= A\textbf{x} + B\textbf{u}\\
    y &= C\textbf{x} + D\textbf{u} \end{aligned} \label{Eq8}
\end{align}
using the following concepts or otherwise the transfer functions modification as presented,   
\begin{figure*}[b!]
\begin{align}
    \dot{\textbf{x}} &= \begin{bmatrix}
    0 & 1 & 0 & 0 \\
    \dfrac{-b_1b_2}{M_1M_2} & 0 & \dfrac{b_1}{M_1}\left(\dfrac{b_1}{M_1} + \dfrac{b_1}{M_2} + \dfrac{b_2}{M_2}\right) - \dfrac{k_1}{M_1} & \dfrac{-b_1}{M_1} \\
    \dfrac{b_2}{M_2} & 0 & -\left(\dfrac{b_1}{M_1} + \dfrac{b_1}{M_2} + \dfrac{b_2}{M_2}\right) & 1 \\
    \dfrac{k_2}{M_2} & 0 & -\left(\dfrac{k_1}{M_1} + \dfrac{k_1}{M_2} + \dfrac{k_2}{M_2}\right) & 0 
    \end{bmatrix}\textbf{x} + \begin{bmatrix}
    0 & 0\\
    \dfrac{1}{M_1} & \dfrac{b_1b_2}{M_1M_2}\\[.75em]
    0 & \dfrac{-b_2}{M_2}\\[.75em]
    \dfrac{1}{M_1} + \dfrac{1}{M_2} & \dfrac{-k_2}{M_2}
    \end{bmatrix}\textbf{u}; \quad y = \begin{bmatrix}
    0 \\
    0 \\
    1 \\
    0 \end{bmatrix}^\top\textbf{x} + \begin{bmatrix}
    0 \\
    0\end{bmatrix}^\top\textbf{u} \label{Eq9}
\end{align}
\end{figure*}
\begin{gather*}
    \int_k\frac{d^kx_n}{dt^k}\;dt = \int_{k-1}\frac{d^{k-1}x_n}{dt^{k-1}}\,dt = x_n,\\
    \frac{1}{M_n}\sum_{i=1}^k F_i = \frac{d^kx_n}{dt^k}, \quad \forall n = 1\to 2; k = 2
\end{gather*}
\newpage
\subsection{The Aircraft Longitudinal-Pitch Dynamics}
The mathematical approaches portraying the aircraft motions with six nonlinear paired are intricate to deal with yet with appropriate assumptions, the schemes of decoupling and linearizing into two axes, lateral and longitudinal perspective, are acceptable. This system focuses on the autonomous aircraft pitch control  
\begin{figure}[h!]
    \centering
    \includegraphics[width=.5\textwidth]{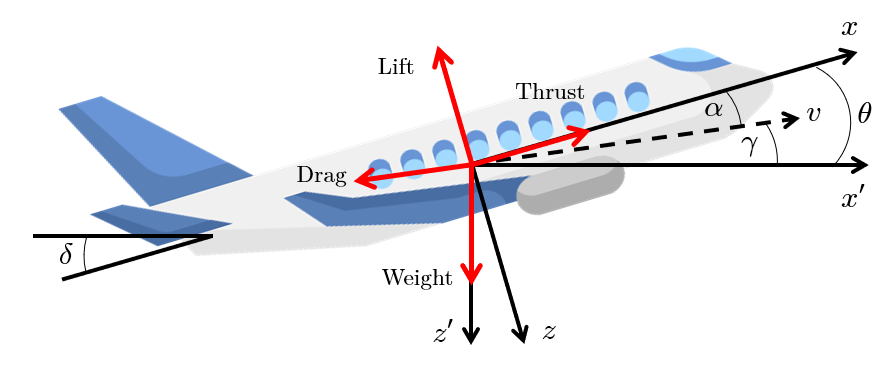}
    \caption{Coordinate dynamics of the aircraft}
    \label{F3}
\end{figure}
being administered by the solely longitudinal axis as shown in Fig.(\ref{F3}). Supposed the steady-cruise occurs in the airplane at certain constants of speed and altitude, then the force variables of drag, lift, weight and thrust systematically balance between the two planes, $x$ and $y$. Keep in mind the more forced assumption of pitch angle alteration, not influencing the velocity, in arbitrary conditions is also applied to simplify the model. This leads to the following longitudinal dynamics with the steady-state variables of attack ($\alpha$) and pitch ($\theta$) angle and pitch rate ($q$) as written in Eq.(\ref{Eq10}), Eq.(\ref{Eq11}), and Eq.(\ref{Eq12}),
\begin{align}
    \dot{\alpha} = \mu\Omega\sigma\left[-\psi_{\alpha_1}\alpha + \psi_{\alpha_2}q - \psi_{\alpha_3}\delta\right] \label{Eq10}
\end{align}
where $\psi_{\alpha_n}, \forall n = 1\to 3$ make of
\begin{align*}
    \psi_{\alpha_1} = \Gamma_\ell + \Gamma_d; \quad \psi_{\alpha_2} = \frac{1}{\mu - \Gamma_\ell}; \quad \psi_{\alpha_3} = \Gamma_w\sin\gamma + \Gamma_\ell
\end{align*}
and the pitch rate ($q$) is written with the following equation of motion
\begin{align}
    \dot{q} = \frac{\mu\Omega}{2I_n}\left[\psi_{q_1}\alpha + \psi_{q_2}q + \psi_{q_3}\delta\right] \quad \longrightarrow \quad \mu = \frac{\rho \textbf{S}\bar{c}}{4m} \label{Eq11}
\end{align}
where $\psi_{q_n}, \forall n = 1\to 3$ constitute
\begin{align*}
    \psi_{q_1} &= \Gamma_m - \eta(\Gamma_\ell + \Gamma_d), &\longrightarrow \eta &= \mu\sigma\Gamma_m\\
    \psi_{q_2} &= \Gamma_m + \sigma\Gamma_m(1 - \mu\Gamma_\ell), &\longrightarrow \sigma &= (1+\mu\Gamma_\ell)^{-1}\\
    \psi_{q_3} &= \eta\Gamma_w\sin\gamma
\end{align*}
The last would be the pitch angle ($\theta$) written as
\begin{align}
    \dot{\theta} = \Omega q \quad\longrightarrow\quad \Omega = \frac{2\textbf{E}_u}{\bar{c}} \label{Eq12}
\end{align}
where $\mu, \Omega, \sigma, \eta$ are the constants being then affected by the following variables; $\delta, \rho, \textbf{S}, \bar{c}, m, \textbf{E}_u, \gamma, I_n$ comprise the deflection angle of elevator, air density, wing area, mean chord length, mass, speed equilibrium, angle of flight trajectory, and the normalized of moment inertia respectively. Beyond that the coefficients are also considered as the coefficient of thrust ($\Gamma_t$), drag ($\Gamma_d$), lift ($\Gamma_\ell$), weight ($\Gamma_w$) and pitch moment ($\Gamma_m$). To obtain the dynamics, it is required to form the state-space from the Laplacian transfer function as in Eq.(\ref{Eq13}) with the respected $c_n, \forall n=1\to 7$, 
\begin{gather}
    sA(s) = c_1A(s) + c_2Q(s) + c_3\Delta(s)\notag\\
    sQ(s) = c_4A(s) + c_5Q(s) + c_6\Delta(s) \label{Eq13}\\
    s\Theta = c_7Q(s) \notag
\end{gather}
and after some algebraic formula, it is achieved this function,
\begin{align}
    \Phi_3(s) = \frac{\Theta(s)}{\Delta(s)} = \frac{1.151s + 0.177}{s^3 + 0.739s^2 + 0.921s} \label{Eq14}
\end{align}
with the standard matrices as in Eq.(\ref{Eq15}) which could be also built from Eq.(\ref{Eq14}), therefore
\begin{align}
    \begin{bmatrix}
    \dot{\alpha}\\[.25em]
    \dot{q}\\[.25em]
    \dot{\theta}
    \end{bmatrix} = \begin{bmatrix}
    c_1 & c_2 & 0\\[.25em]
    c_4 & c_5 & 0\\[.25em]
    0 & c_7 & 0
    \end{bmatrix}\begin{bmatrix}
    \alpha\\[.25em]
    q\\[.25em]
    \theta
    \end{bmatrix} + \begin{bmatrix}
    c_3\\[.25em]
    c_6\\[.25em]
    0
    \end{bmatrix}\delta; \qquad y = \theta \label{Eq15}
\end{align}

\subsection{DC Motor Systems - Speed \& Position Perspectives}
\begin{figure}[h!]
    \centering
    \includegraphics[width=.445\textwidth]{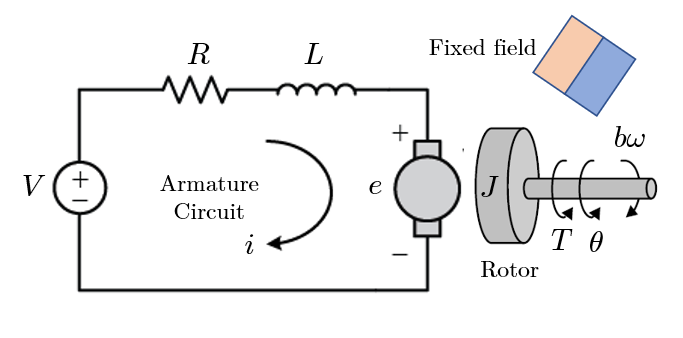}
    \caption{Free-body rotor design}
    \label{F4}
\end{figure}

The last dynamical system would be one of the most common actuators being used in electrical drives, which is focused on the two measured variables of speed and position. This system illustrates the translational rotating rotor-motion paired with the wheels as shown in Fig.(\ref{F4}). The rotor plant is supposed to have the voltage ($V$) input working on the armature of the motor whereas the outputs capture the two states, the position $\theta$ and the speed $\dot{\theta}$ of the shaft. Furthermore, the two objects associated to the input-output are assumed 
to be rigid while regarding the resistive-contacting force, the model of the friction torque linearly parallels to angular velocity. With the steady value assumption of magnetic field, the torque, having the positive linear combination with the current and the magnetic field, is then solely corresponding to the current ($i$) with certain torque parameter ($\kappa_t$), such that  
\begin{align}
    T = \kappa_ti \label{Eq16}
\end{align}
where the opposite emf ($e$) is positively affected by the multiplication between certain electromotive force parameter ($\kappa_e$) which equals to ($\kappa_t$) and the shaft speed $\dot{\theta}$ 
\begin{align}
    e = \kappa_e\dot{\theta} \label{Eq17}
\end{align}
and from Eq.(\ref{Eq16}) and Eq.(\ref{Eq17}), the respected models in terms of the force Newton's law and the voltage Kirchhoff's could be written in Eq.(\ref{Eq18})
\begin{align}\begin{aligned}
    T\ddot{\theta} + b\theta &= \kappa i\\
    L\frac{di}{dt} + Ri &= V - \kappa\dot{\theta} \end{aligned} \label{Eq18}
\end{align}
which are then constructed by the Laplacian functions stated as
\begin{align}\begin{aligned}
    s(Js + b)\Theta(s) &= \kappa I(s)\\
    (Ls + R)I(s) &= V(s) - \kappa s\Theta(s)\end{aligned} \label{Eq19}
\end{align}
From Eq.(\ref{Eq19}), the term of ($s$) in the position $\Theta(s)$ is then mixed as the output velocity whereas the $I(s)$ becomes the equality variable between the two equations, making the input voltage $V(s)$ as written in Eq.(\ref{Eq20}), therefore
\begin{align}
    \Phi_4(s) = \frac{\dot{\Theta}(s)}{V(s)} = \frac{\kappa}{(Js + b)(Ls + R) + \kappa^2} \label{Eq20}
\end{align}
Either the Laplacian term in Eq.(\ref{Eq20}) or by constructing from the equation models could be easily turned into state-space representation with two measured variables of position and speed as denoted in Eq.(\ref{Eq21}), 
\begin{align}\begin{aligned}
    \frac{d}{dt}\begin{bmatrix}
    \theta\\[.25em]
    \dot{\theta}\\[.25em]
    i
    \end{bmatrix} &= \begin{bmatrix}
    0 & 1 & 0\\[.25em]
    0 & \frac{-b}{j} & \frac{\kappa}{j}\\[.25em]
    0 & \frac{-\kappa}{L} & \frac{-R}{j}
    \end{bmatrix}\begin{bmatrix}
    \theta\\[.25em]
    \theta\\[.25em]
    i
    \end{bmatrix} + \begin{bmatrix}
    0\\[.25em]
    0\\[.25em]
    \frac{1}{L}\end{bmatrix}V\\
    y &= \begin{bmatrix}
    1 & 0 & 0\\[.25em]
    0 & 1 & 0
    \end{bmatrix}\textbf{x} \longrightarrow  \begin{bmatrix}
    \theta\\[.25em]
    \dot{\theta}\end{bmatrix}\end{aligned} \label{Eq21}
\end{align}

\section{Control Designs}\label{Sec3}
\begin{figure}[h!]
    \centering
    \includegraphics[width=.5\textwidth]{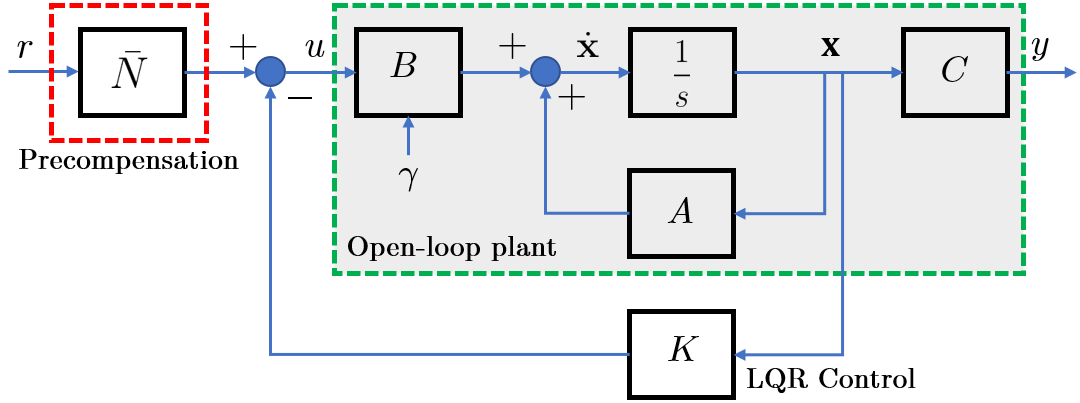}
    \caption{A precompensator state-feedback system}
    \label{F5}
\end{figure}
The Control method used is the feedback aided with pre-compensation along with some disturbance under some desired \newpage
\noindent condition. Fig.(\ref{F5}) explains the control scenario of the feedback aided compensation $\bar{N}$ where it requires standard pole placement to obtain $K$. Regarding the compensation, it has two initial setup of $s_1$ as the length of matrix $A$ and $s_2$ defining the ones vector of output in addition to zeros vector of $s_1$, such that, 
\begin{align*}
    s_1 = \textrm{length}(A) \qquad s_2 = [\textrm{zeros}([1,s_1]),\; 1]
\end{align*}
where the compensation ($\bar{N}$) is constructed using the formulas
\begin{gather*}
    N = \textrm{inv}\left(A,B; C,D\right) \times s_2^\top \\
    \bar{N} = N_u + KN_x \longrightarrow Nx = N(1:s_1) \textrm{ and } N_u = N(1+s_1)
\end{gather*}
which could vanish the error. Furthermore, as regards the first plant, the cruise is designed to be constant in any two different points for 50s. The first 30s is for 10 m/s as the reference whereas the rest 20s is to maintain 7 m/s with ($-1.5$) pole for finding the gain $K$. The second plant requires the disturbance $\gamma$ in addition to input signal $u$ and from this, the control should stabilize the output $y_1$ in under 5s. Keep in that while having two inputs ($u,\gamma$), the state-feedback could only control the input signal of the first column $B(\textbf{1})$, such that
\begin{align*}
    \dot{x} = (A - B(\textbf{1})\times K) + B[U,\Gamma]^\top
\end{align*}
where the characteristic polynomial of the system is then written as the $\det[sI - (A-B(\textbf{1})K)]$ instead of the standard $\det[sI - (A-BK)]$ along with the minimizing compensation $\bar{N}$ algorithm. With respect to the third plant, the full-rank ($n$) analysis of controllable and observable is done, ensuring to place the poles around complex $s-$plane. The input $u$ would be the desired pitch angle $\theta_r$ and the $K$ times the full-state measured $\textbf{x}$ however, since this is the higher-order dynamics, the more advanced technique is applied with the weighted matrices of $R$ and modified $Q$ applying the varied-constant $p$ and matrix $C$. After that, the gain $K$ and the pre-compensation are ready to be implemented. Finally, the rotor dynamics have poles in sequence of speed $v$ ($-5\pm i$) and position $\theta$ ($-100\pm 100i, -200$) along with the same scenarios of finding the gain $K$ and the minimizing compensation $\bar{N}$. Beyond that, the stability analysis are discussed further to guarantee the desired outputs along with the proposed estimation mechanism.

\section{Stability Analysis}\label{Sec4}
\begin{figure*}
    \centering
    \includegraphics[width=\textwidth]{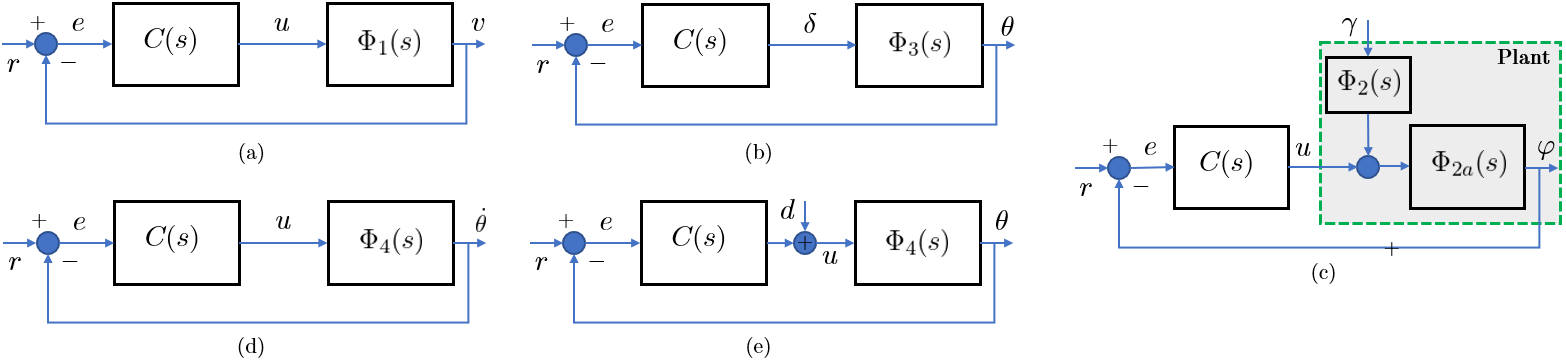}
    \caption{The block diagrams of the tested plants according to the transfer function $\Phi_n, \forall n = 1\to 4$ as the basis of stability analysis}
    \label{C}
\end{figure*}
The design of the control systems $C(s)$ for each plant is shown in Fig.(\ref{C}) according to certain transfer function $\Phi_n, \forall n = 1\to 4$ with $(r,e,d)$ represent the reference, error and disturbance. For PID-typed control, the signal control would be evaluated through Eq.(\ref{Eq22}),
\begin{align}
    u(t) = K_pe(t) + K_i\int e(t)dy + K_d\frac{de}{dt} \label{Eq22}
\end{align}
while the control transfer functions in Eq.(\ref{Eq23}) would be adjusted to the plant $\Phi_n$ based on Eq.(\ref{Eq3}), Eq.(\ref{Eq7}), Eq.(\ref{Eq14}), and Eq.(\ref{Eq20}), especially when finding the position in $\Phi_4(s)$ it is then required to integrate, therefore
\begin{figure*}
     \centering
     \begin{subfigure}[b]{0.325\textwidth}
         \centering
         \includegraphics[width=\textwidth]{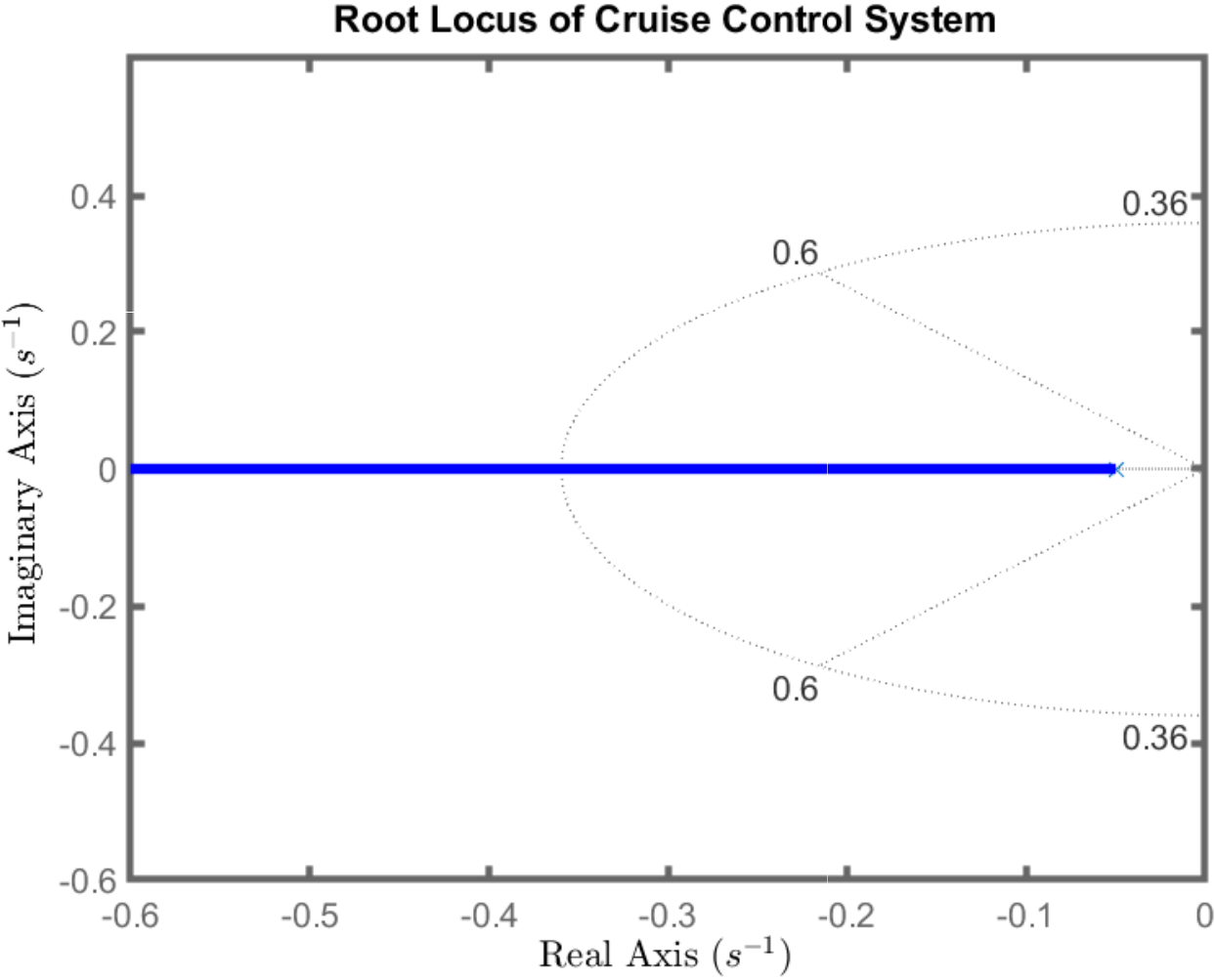}
         \caption{}
         \label{RL11}
     \end{subfigure}
     \begin{subfigure}[b]{0.325\textwidth}
         \centering
         \includegraphics[width=\textwidth]{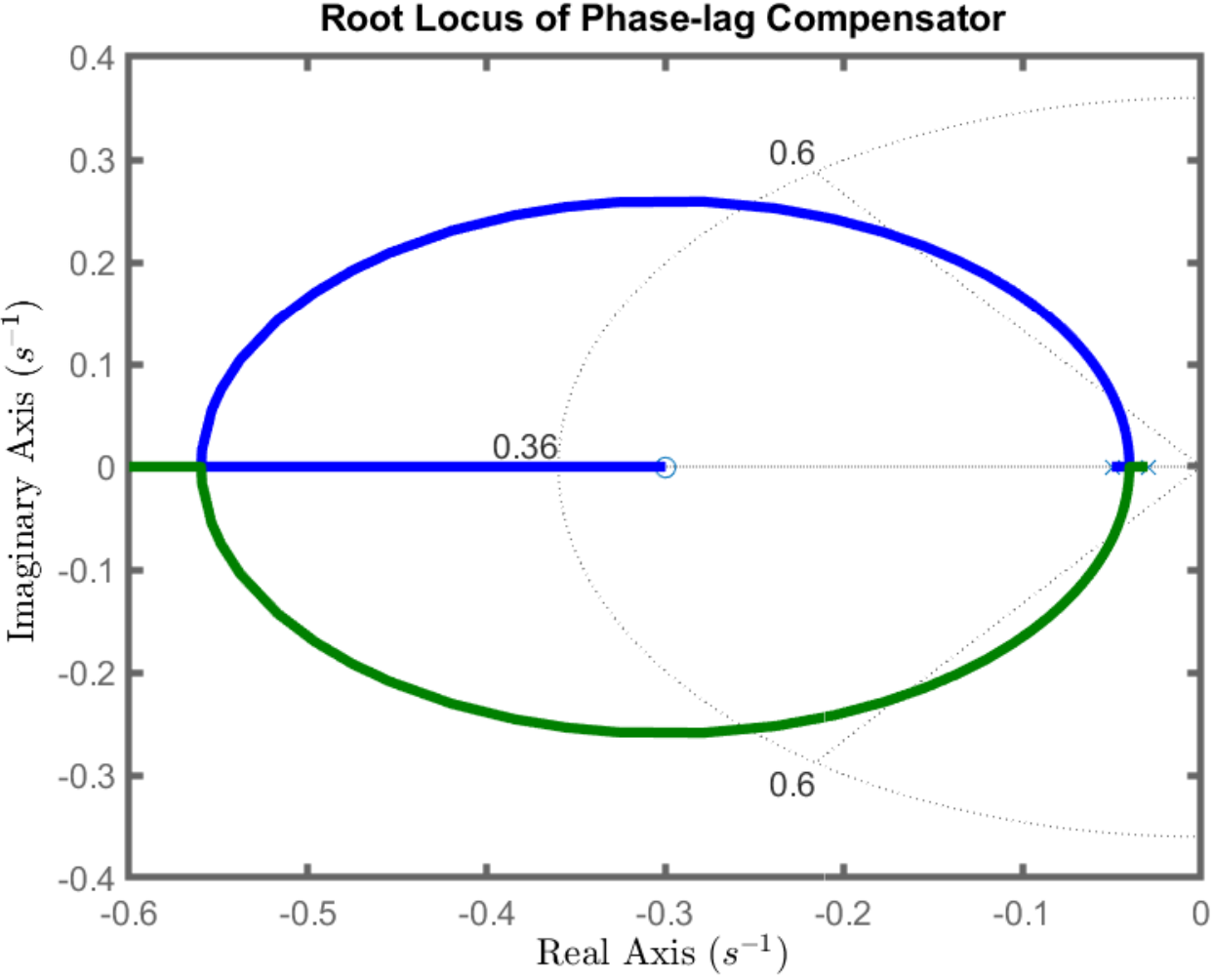}
         \caption{}
         \label{RL12}
     \end{subfigure}
     \begin{subfigure}[b]{0.33\textwidth}
         \centering
         \includegraphics[width=\textwidth]{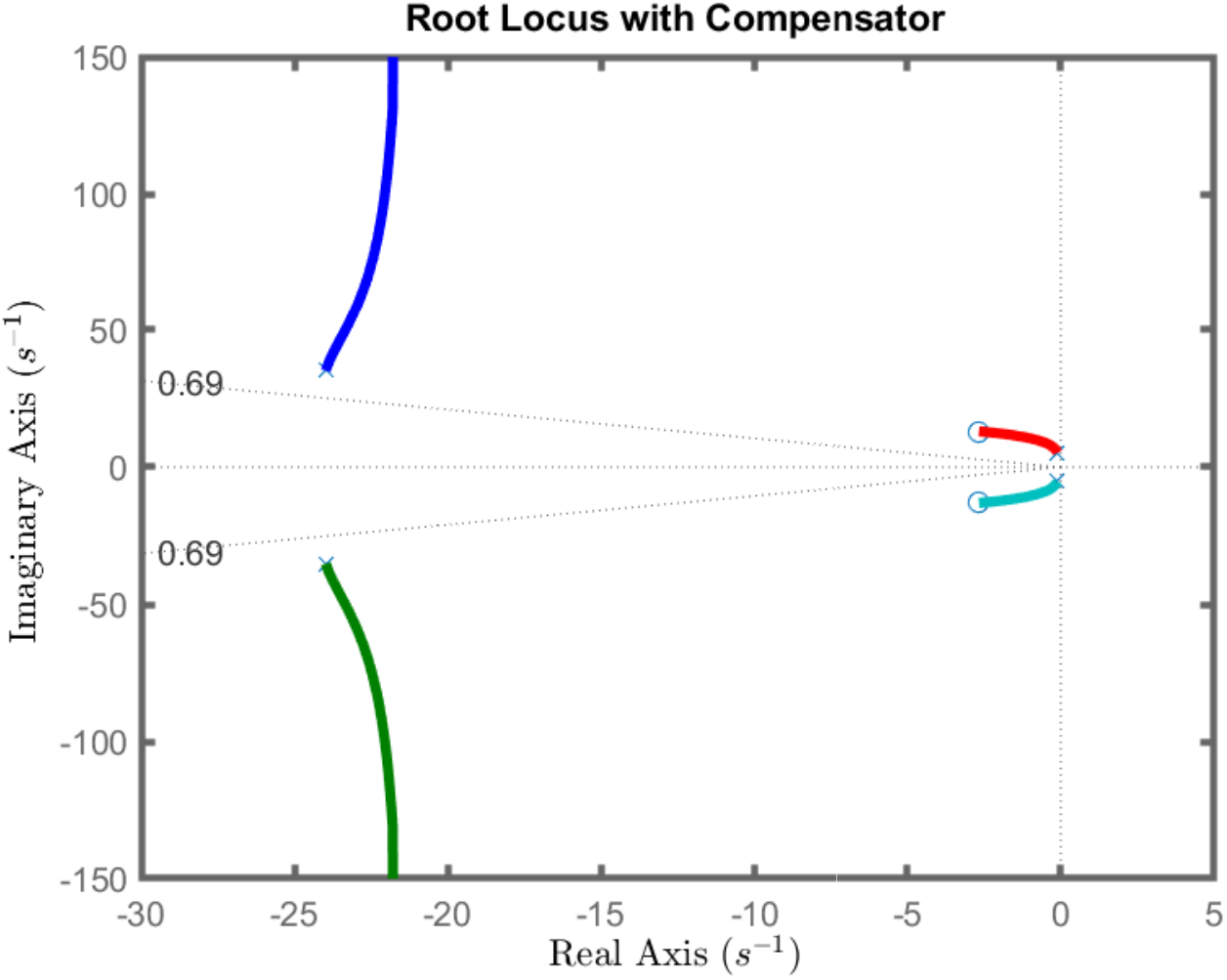}
         \caption{}
         \label{RL21}
     \end{subfigure}\\[.25em]
     \begin{subfigure}[b]{0.33\textwidth}
         \centering
         \includegraphics[width=\textwidth]{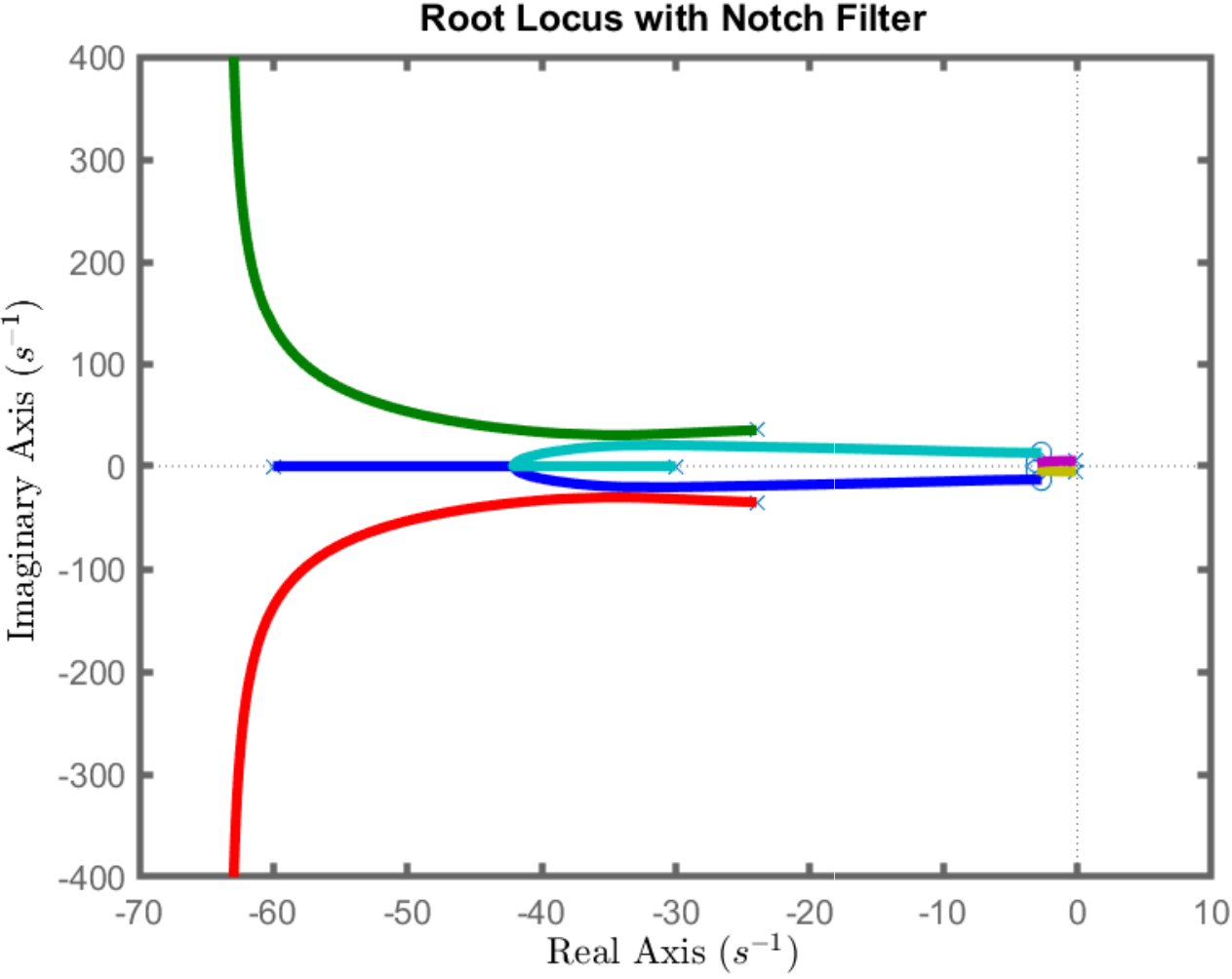}
         \caption{}
         \label{RL11}
     \end{subfigure}
     \begin{subfigure}[b]{0.325\textwidth}
         \centering
         \includegraphics[width=\textwidth]{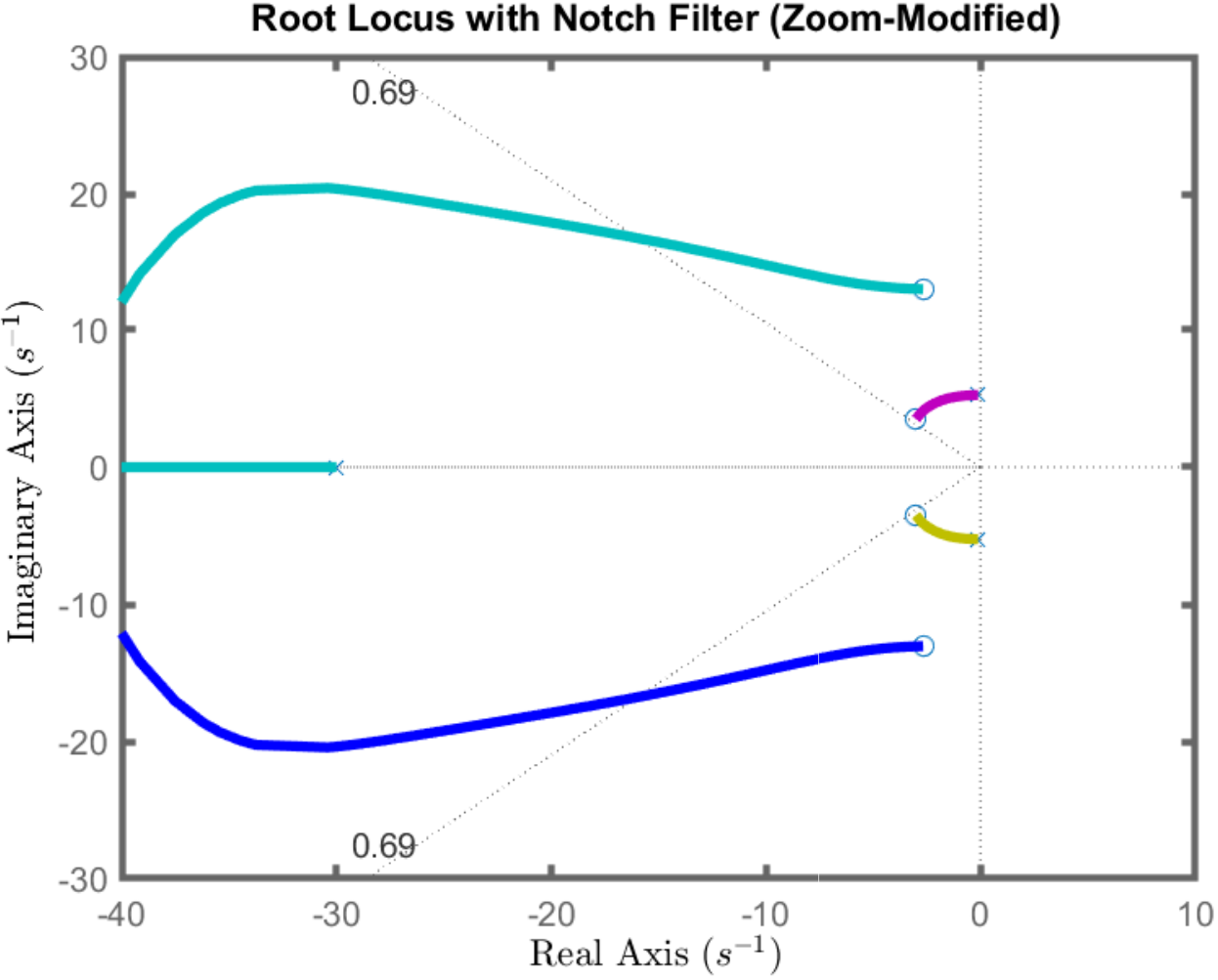}
         \caption{}
         \label{RL12}
     \end{subfigure}
     \begin{subfigure}[b]{0.325\textwidth}
         \centering
         \includegraphics[width=\textwidth]{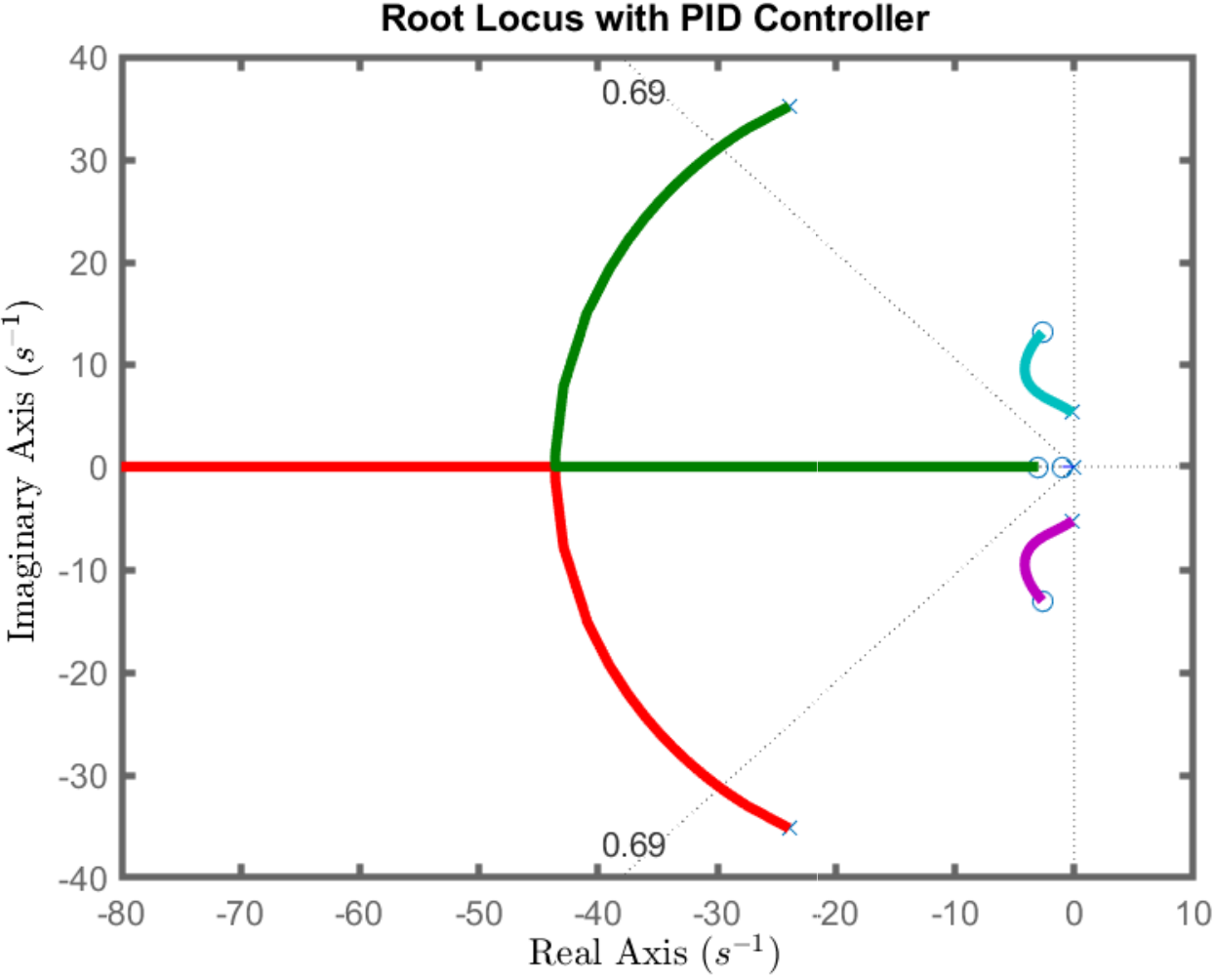}
         \caption{}
         \label{RL21}
     \end{subfigure}\\[.25em]
     \begin{subfigure}[b]{0.325\textwidth}
         \centering
         \includegraphics[width=\textwidth]{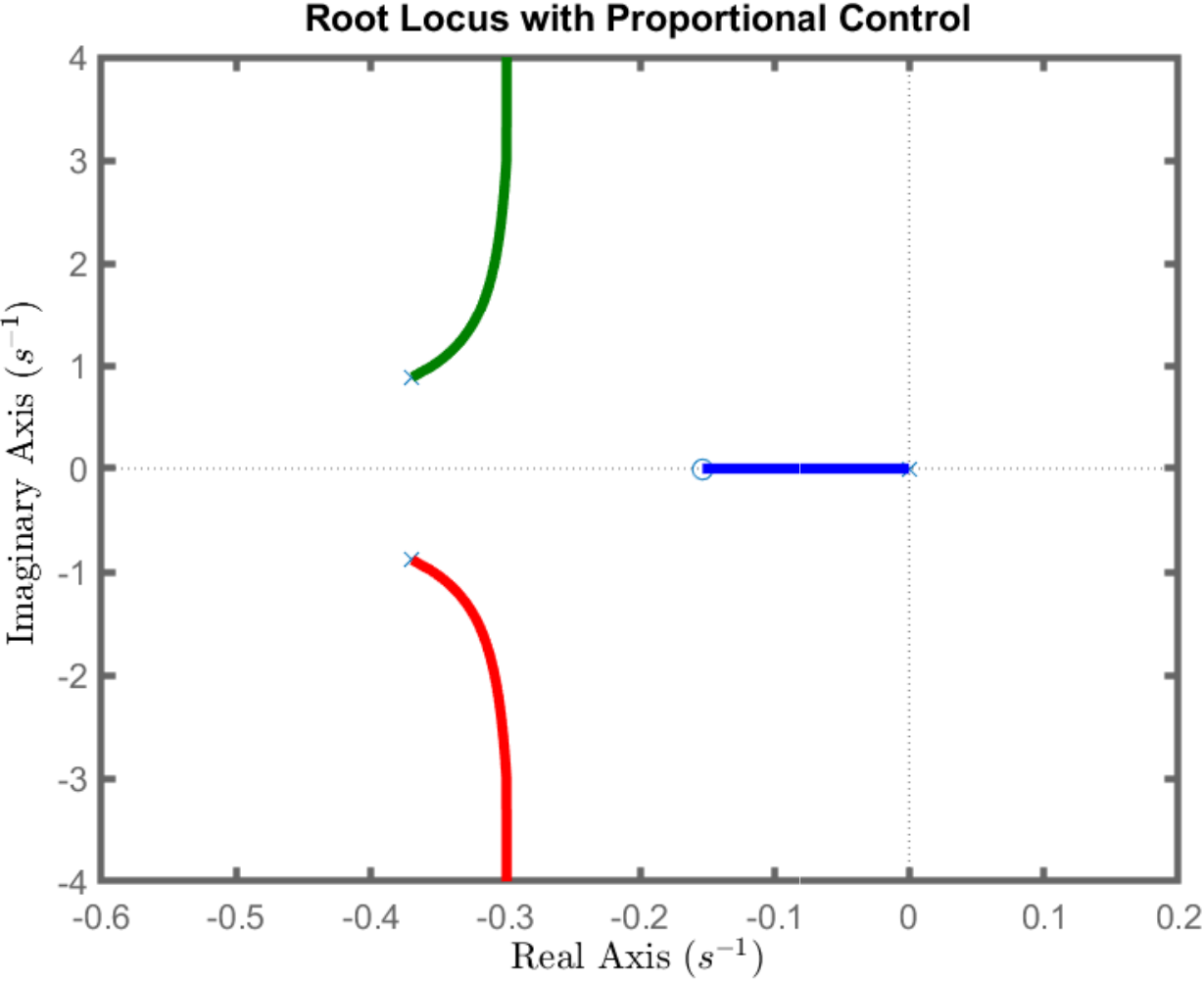}
         \caption{}
         \label{RL11}
     \end{subfigure}
     \begin{subfigure}[b]{0.325\textwidth}
         \centering
         \includegraphics[width=\textwidth]{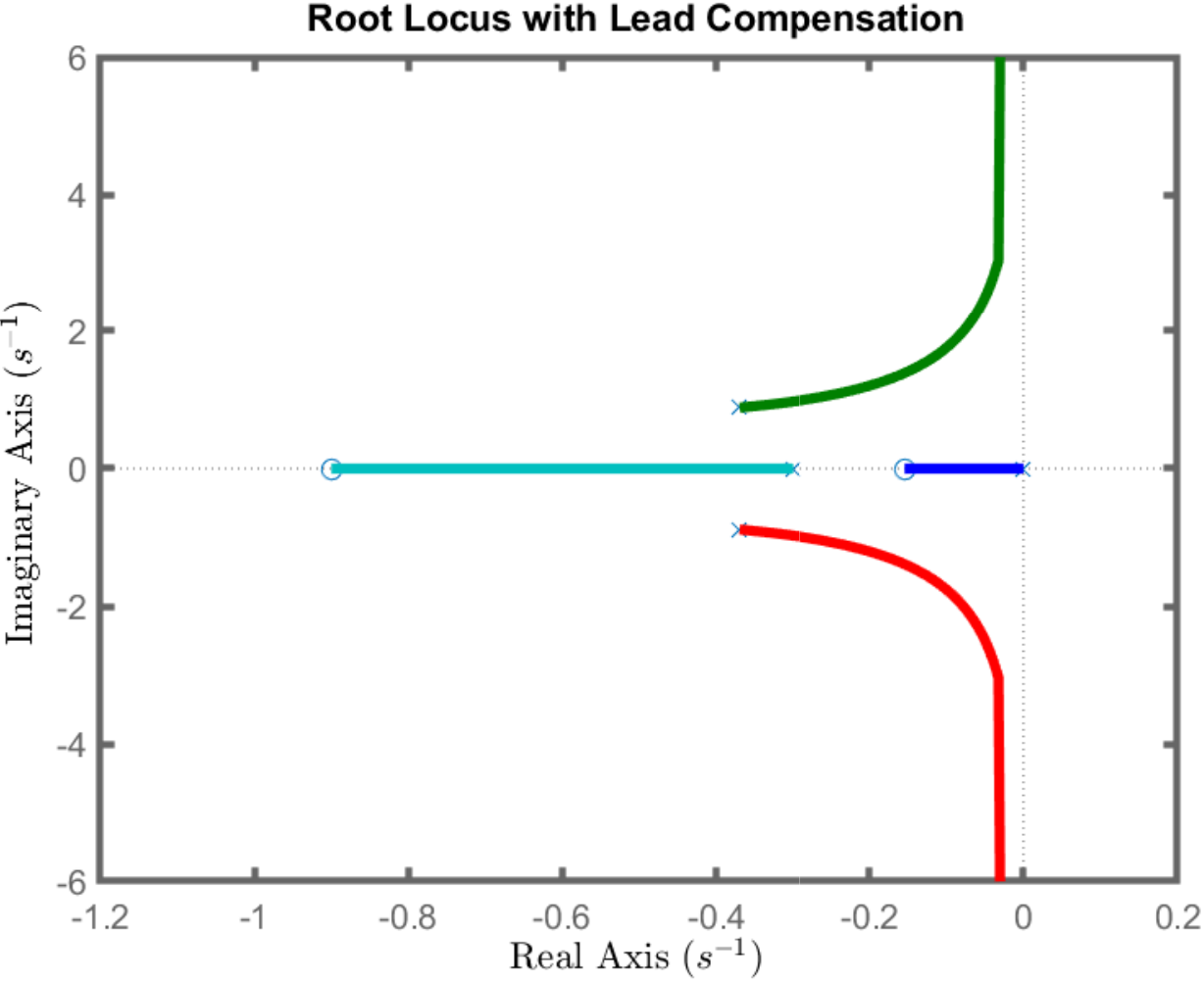}
         \caption{}
         \label{RL12}
     \end{subfigure}
     \begin{subfigure}[b]{0.325\textwidth}
         \centering
         \includegraphics[width=\textwidth]{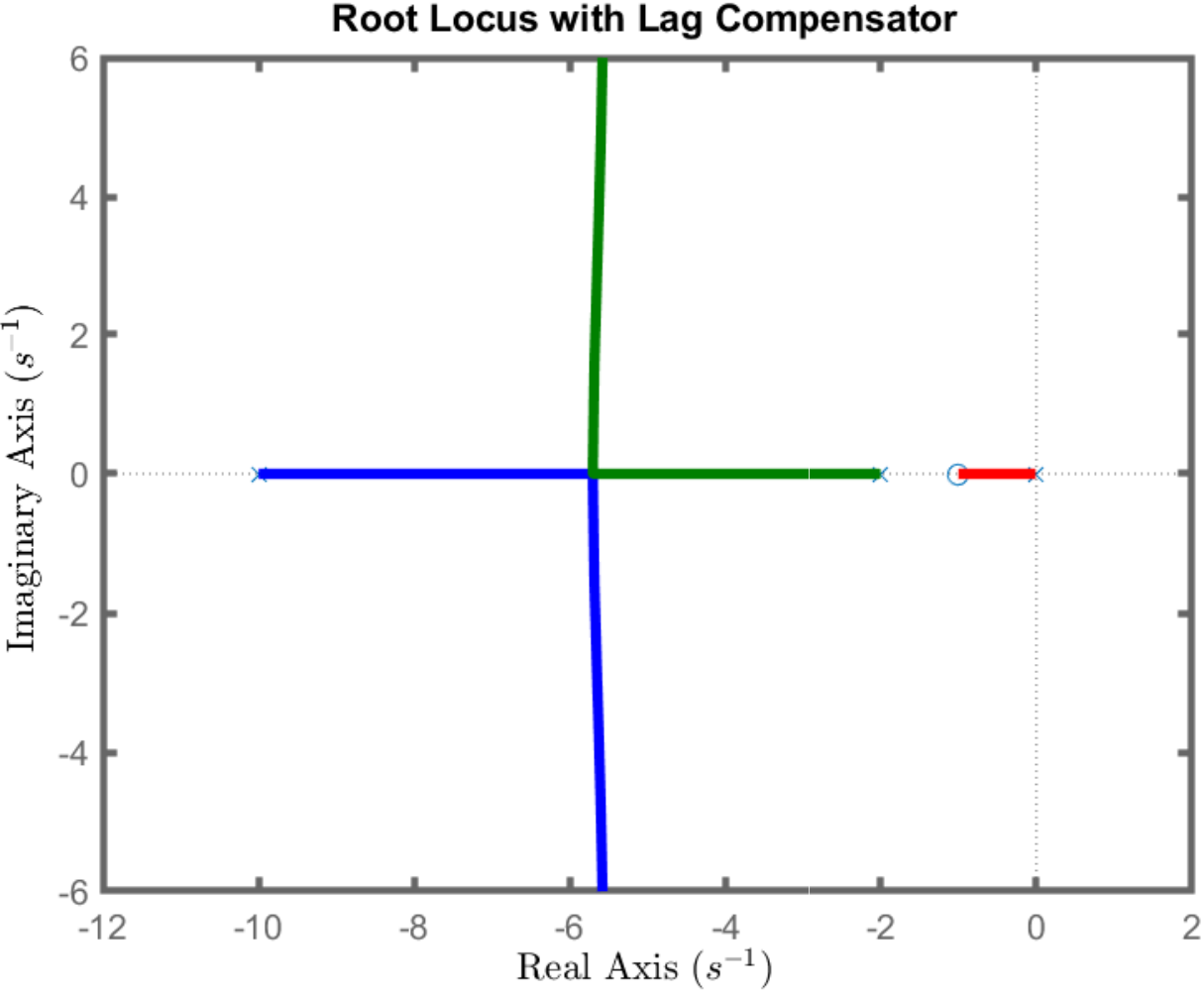}
         \caption{}
         \label{RL21}
     \end{subfigure}\\[.25em]
     \begin{subfigure}[b]{0.3175\textwidth}
         \centering
         \includegraphics[width=\textwidth]{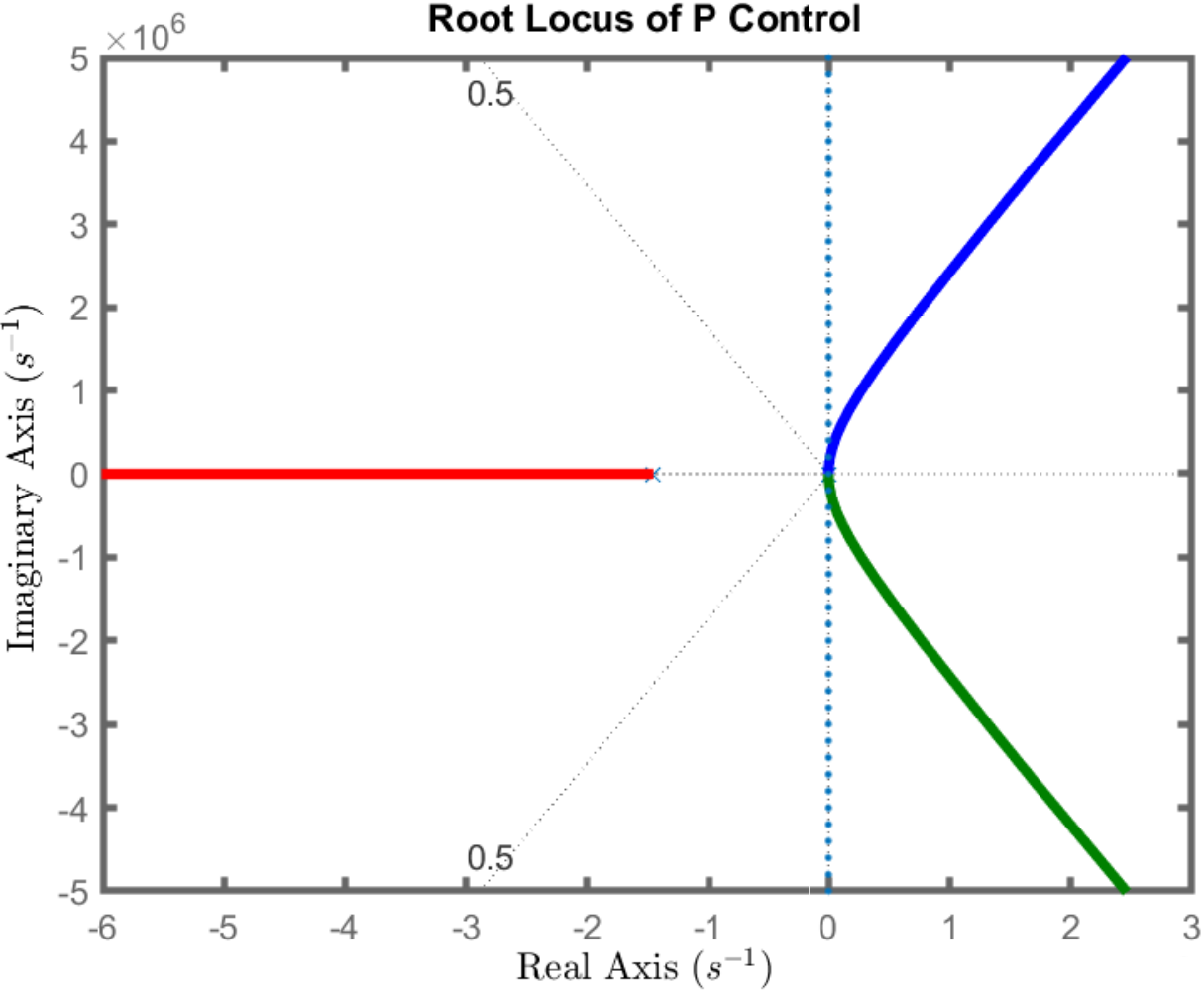}
         \caption{}
         \label{RL11}
     \end{subfigure}
     \begin{subfigure}[b]{0.33\textwidth}
         \centering
         \includegraphics[width=\textwidth]{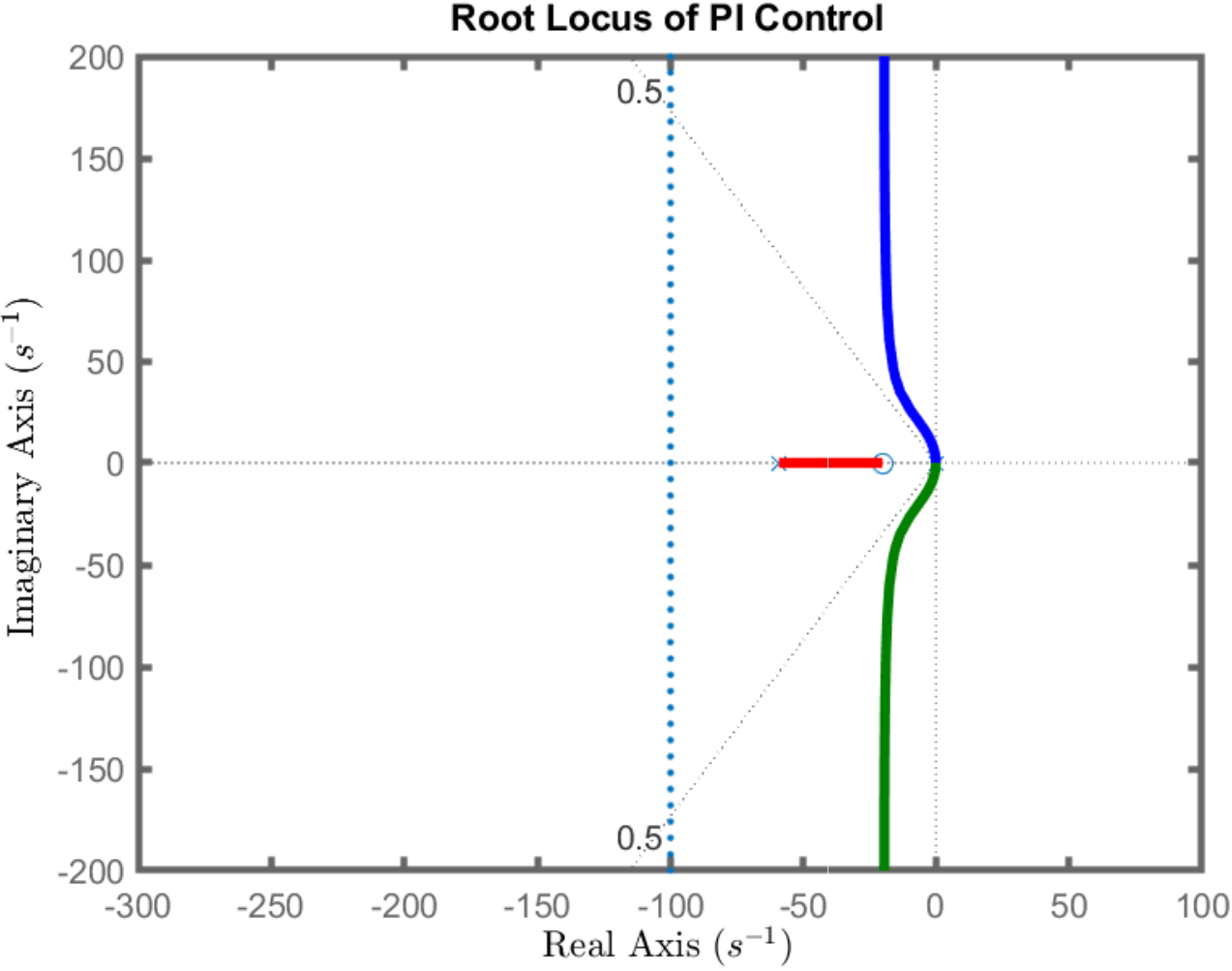}
         \caption{}
         \label{RL12}
     \end{subfigure}
     \begin{subfigure}[b]{0.33\textwidth}
         \centering
         \includegraphics[width=\textwidth]{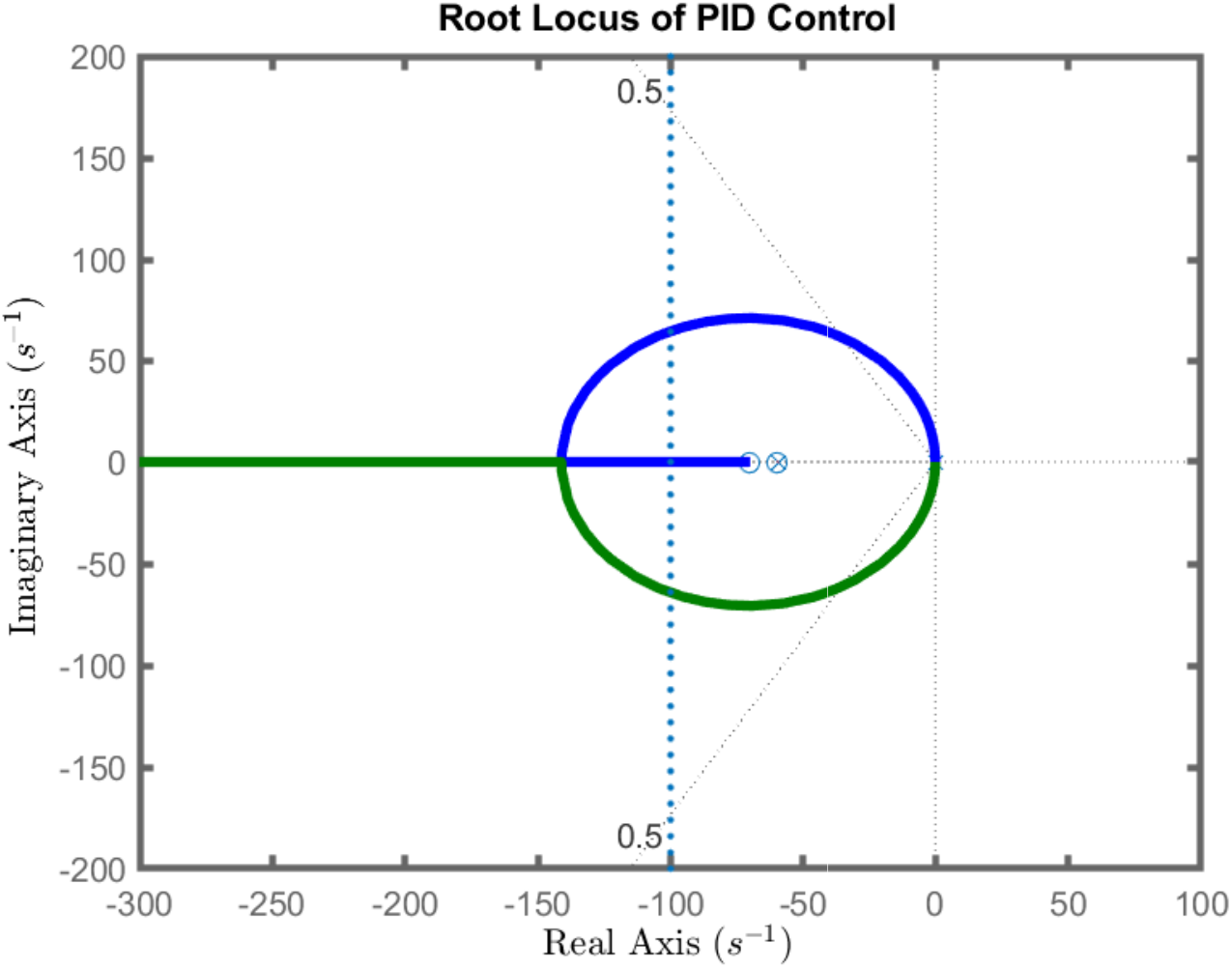}
         \caption{}
         \label{RL21}
     \end{subfigure}\\
     \caption{The root-locus analysis of: 1) the first dynamics $\Phi_1$ with proportional (a) and lag-controller (b); 2) the second plant $\Phi_{2\alpha}$ (c) with notch filter (d) and its magnifier (e) to the limit along with PID-control (f); 3) the third system $\Phi_3$ with original-induced by proportional control (g) and the modification of lead-compensation scenario (h); 4) while the last transfer functions comprise the speed $\Phi_4$ with lag-control (i) and the position $\Phi_4$ with proportional (j), the PI control (k) and the PID-control (l) }
     \label{RL}
\end{figure*}
\begin{align}
    C(s) = \begin{cases}
    K_pE(s), &\rightarrow \textrm{P}\\[.75em]
    (K_p + K_ds)E(s), &\rightarrow \textrm{PD}\\[.5em]
    \dfrac{K_P s + K_i}{s}E(s), &\rightarrow \textrm{PI}\\[.5em]
    \dfrac{K_ds^2 + K_Ps + K_i}{s}E(s), &\rightarrow \textrm{PID}
    \end{cases} \label{Eq23}
\end{align}
Moreover, the damping ratio ($\zeta$) and natural frequency ($\omega_n$) for each plant should be well-defined using Eq.(\ref{Eq24}), such that 
\begin{align}
    \omega_n \geq \frac{1.8}{t_r}, \qquad \zeta\geq\sqrt{\dfrac{\ln^2(M_p)}{\pi^2 + \ln^2(M_p)}} \label{Eq24}
\end{align}
Keep in mind that other than PID combination in Eq.(\ref{Eq23}), there are lag-control and lead-compensation being used in analysing the stability. For instance the lag-control of the first dynamic $\Phi_1$ as in Eq.(\ref{Eq25}),
\begin{align}
    C(s) = \frac{s + z_0}{s + p_0} \longrightarrow \frac{V(s)}{R(s)} = \frac{s + z_0}{ms^2 + (b + mp_0)s + bp_0} \label{Eq25}
\end{align}
where the closed-loop design constitutes in Eq.(\ref{Eq26}),
\begin{align}
    \frac{V(s)}{U(s)} = \frac{K_p(s + z_0)}{ms^2 + (b + mp_0 + K_p)s + (bp_0 + K_pz_0)} \label{Eq26}
\end{align}
and the led mechanism could be seen with another additional gain $K_\ell$ in $C(s)$ and putting the value of zero $z_0$ less than that of pole $p_0$. While the lag considers the right of $s-$ plane, the counterpart led places on the left side with results in Fig.(\ref{RL}). 

\section{The Modified Unscented Filtering}\label{Sec5}
The common algorithm of Unscented Kalman Filter covers the vital response for the standard-EKF question in terms of the best-opted Gaussian random variables (GRV). It gives the broader deterministic sampled-values and this has captured the posterior mean along with its covariance of the opted GRV when being matched to the more dynamic non-linear systems,
\begin{align}\begin{aligned}
    \textbf{x}_{k+1} &= f(\textbf{x}_k, \textbf{u}_k) + q_k \\ \textbf{y}_k &= h(\textbf{x}_k) + r_k
    \end{aligned} \label{Eq27}
\end{align}
where the matrices of $\textbf{x}_k, \textbf{u}_k, f_k, h_k, w_k, v_k$ and $\textbf{y}_k$ comprise state, control signal, system and measurement model, process and measurement noise, and measurement vector respectively. Considering to input a state-variable ($\textbf{x}$), with the properties of $\bar{\textbf{x}}$ and $\textbf{P}_x$ with $L-$dimension, into the non-linear model $f(\bullet)$ to gain the measured $\textbf{y}_k$, the modified matrix $\chi_i$ with $\forall i = 0 \to (2L + 1)$ \textit{sigma} values, such that 
\begin{align}\begin{cases}
    \chi_0 = \bar{\textbf{x}} &i=0\\
    \chi_i = \bar{\textbf{x}} + \left(\sqrt{\left(L + \lambda\right)\textbf{P}_{k-1}}\right)_i &i=1,\dots,L\\
    \chi_i = \bar{\textbf{x}} - \left(\sqrt{\left(L + \lambda\right)\textbf{P}_{k-1}}\right)_i &i=L+1,\dots,2L
\end{cases}\label{Eq28}
\end{align}
in which $\lambda = \alpha^2(L+\kappa) - L$ explains the weighted value and a parameter ($\alpha$) as the key factor defines the distribution of the sigma values $(\chi_i)$ surrounding the expected value ($\bar{\textbf{x}}$) with the value in a range of $10^{-4}$ and 1. While ($\kappa$) constitutes the second parameter, either 0 or $(3-L)$, the third ($\beta$), demonstrating the previous data of the state, makes of the exact 2 as the optimal. Moreover, the method of the standard UKF is written and to differ from the preceding discussion, this applies the $(n)-$th iteration in place of ($k$),
\begin{enumerate}
    \item \textbf{Initialization} setup the inputs $\hat{\textbf{x}}_0$ and $\textbf{P}_0$
    \item For $n = 1\to\infty$, iterate the algorithms from Eq.(\ref{EqA1}) to Eq.(\ref{EqA11}). First, calculate the sigma values ($\chi$),  
\begin{align}
    \chi_{a,n-1}^{(i)} &= \begin{bmatrix}\textbf{x}_{a,n-1}^{(i)} & \textbf{x}_{a,n-1}^{(i)} \pm \sqrt{(L + \lambda)\textbf{P}_{a,n-1}} 
    \end{bmatrix}\tag{$\alpha_1$}\label{EqA1}
    \intertext{\item \textbf{Time updated:} Compute the sigma $\chi_{a,n-1}^{(i)}$ using the function system to obtain $\hat{\chi}_{a,n}^{(i)}$ and $\hat{\bf{y}}_{n}^{(i)}$. Furthermore, solving the values of $\hat{\textbf{P}}_{a,n}$, $\hat{\bf{y}}_{n}$, and $\hat{\textbf{P}}_{a,n}$, implementing the following formulas}
    \hat{\chi}_{a,n}^{(i)} &= \textbf{f}\left(\chi_{a,n-1}^{(i)}, \textbf{u}_n\right)\tag{$\alpha_2$}\\
    \hat{\textbf{x}}_{a,n} &= \sum_{i=0}^{2L} W^{(i)} \hat{\chi}_{a,n}^{(i)}\tag{$\alpha_3$}\\
    \hat{\textbf{P}}_{a,n} &= \sum_{i=0}^{2L} W^{(i)}\left[\hat{\chi}_{a,n}^{(i)} - \hat{\textbf{x}}_{a,n}\right]\left[\hat{\chi}_{a,n}^{(i)} - \hat{\textbf{x}}_{a,n}\right]^\top\tag{$\alpha_4$}\label{EqA4}\\
    \hat{\bf{y}}_{n}^{(i)} &= \textbf{g}\left(\hat{\chi}_{a,n}^{(i)}, \textbf{u}_n\right)\tag{$\alpha_5$}\\
    \hat{\bf{y}}_{n} &= \sum_{i=0}^{2L} W^{(i)}  \hat{\bf{y}}_{n}^{(i)}\tag{$\alpha_6$}
    \intertext{\item \textbf{Measurement updated:} Compute the variables of $\hat{\textbf{S}}_{n}$, $\hat{\textbf{K}}_{n}^{xy}$, $\textbf{W}_n$ along with the updated states $\hat{\textbf{x}}_n^+$ and error covariance $\textbf{P}_n$,}
    \hat{\textbf{S}}_{n} &= \sum_{i=0}^{2L} W^{(i)}\left[\hat{\bf{y}}_{n}^{(i)} - \hat{\bf{y}}_{n}\right]\left[\hat{\bf{y}}_{n}^{(i)} - \hat{\bf{y}}_{n}\right]^\top\tag{$\alpha_7$}\\
    \hat{\textbf{K}}_{n}^{xy} &= \sum_{i=0}^{2L} W^{(i)}\left[\hat{\chi}_{n}^{(i)} - \hat{\textbf{x}}_{n}\right]\left[\hat{\bf{y}}_{n}^{(i)} - \hat{\textbf{x}}_{n}\right]^\top\tag{$\alpha_8$}\\
    \textbf{W}_n &= \hat{\textbf{K}}_{n}^{xy}\hat{\textbf{S}}_{n}^{-1}\tag{$\alpha_9$}\\
    \hat{\textbf{x}}_n^+ &= \hat{\textbf{x}}_n + \textbf{W}_n z_n \longrightarrow z_n = \textbf{y}_n - \hat{\textbf{y}}_n^{(i)}\tag{$\alpha_{10}$}\\
    \textbf{P}_n &= \hat{\textbf{P}}_{n} - \textbf{W}_n\hat{\textbf{S}}_{n}\textbf{W}_n^\top \tag{$\alpha_{11}$}\label{EqA11}
\end{align}
with the magnitudes of $W_i$ are stated as follows,
\begin{figure}
    \centering
    \includegraphics[width=.5\textwidth]{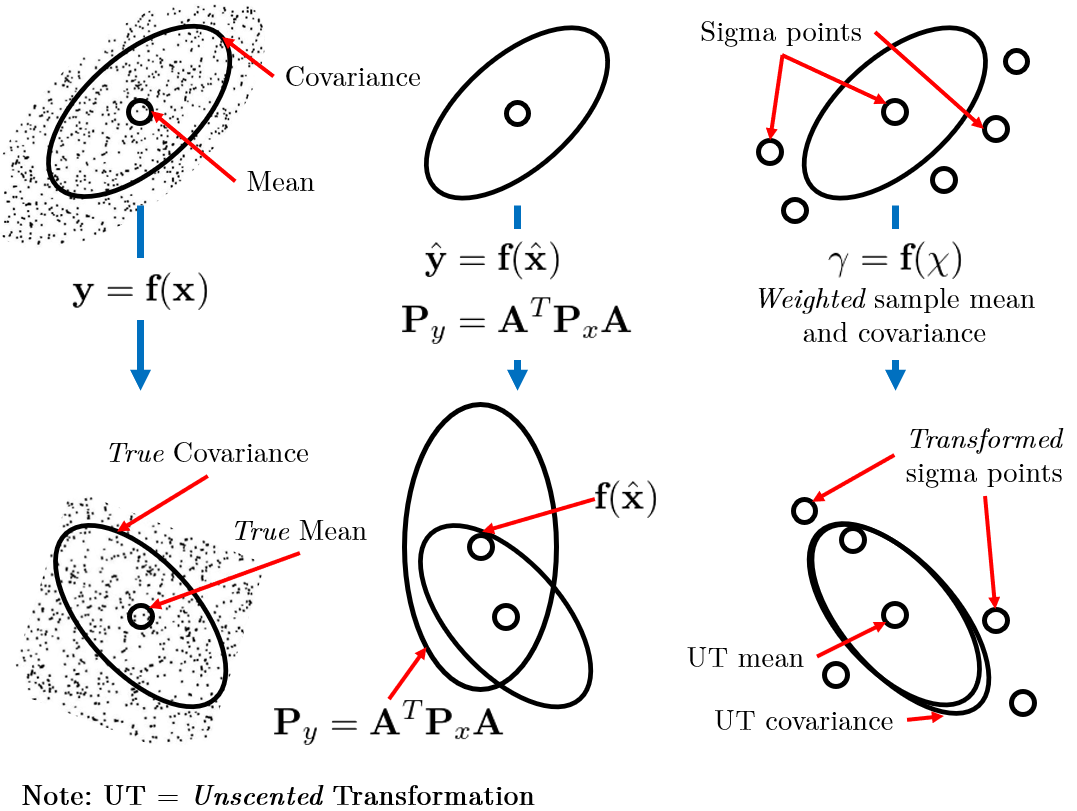}
    \caption{The UT propagation of mean and covariance: (\textit{left}) The actual; (\textit{center}) First-linearization of EKF; (\textit{right}) The UT itself}
    \label{F8}
\end{figure}
\begin{align}\begin{aligned}
    W_0^{(m)} &= \frac{\lambda}{L + \lambda} \\ W_0^{(c)} &= \frac{\lambda}{L + \lambda} + \left(1-\alpha^2 + \beta\right) \\ W_i^{(m)} &= W_i^{(c)} = \frac{1}{2(L + \lambda)}, \forall i = 1\to 2L 
\end{aligned}\label{Eq29}
\end{align}
\end{enumerate}
The weighted average comes up inspired by the consensus-based algorithms with some matrix modification of the pseudo measurement as written in \cite{R31}. Nevertheless this was linear approximation of the pseudo matrix \cite{R32} and according to \cite{R33} and \cite{R34}, the consensus was applied by \cite{R30} without pseudo matrix. More specifically, each node ($i$) calculates locally with the weighted average between the estimated state and error covariance ($\hat{\textbf{x}}_n^i,\textbf{P}_n^i$) in neighborhood region $\mathcal{N}_i$ with certain magnitude of $\pi^{(1,j)}, j\in\mathcal{N}_i$. The coupled $(\hat{\textbf{x}}_n^i,\textbf{P}_n^i)_{i\in\mathcal{N}}$ is then supposed to be weighted average if $\ell\to\infty$ as in Eq.(\ref{Eq30}),  
\begin{align}
    \left(\hat{\textbf{x}}_n^i,\textbf{P}_n^i\right)_{i\in\mathcal{N}} \xrightarrow[]{\textrm{weighted}}
    \left(\hat{\textbf{x}}_n^\ast, \textbf{P}_n^\ast\right) = \lim_{\ell\to\infty}\left(\hat{\textbf{x}}_{n,\ell}^i, \textbf{P}_{n,\ell}^i\right) \label{Eq30}
\end{align}
where the coupled-term $(\hat{\textbf{x}}_{n,\ell}^i,\textbf{P}_{n,\ell}^i)_{i\in\mathcal{N}}$ comprises the data provided at ($i$) point at the $\ell-$th cycle, satisfying Eq.(\ref{Eq31})
\begin{align}\begin{aligned}
    \hat{\textbf{x}}_{n,\ell+1}^i &= \sum_{j\in\mathcal{N}_i}\pi^{(ij)}\hat{\textbf{x}}_{n,\ell}^j \\ \textbf{P}_{n,\ell+1}^i &= \sum_{j\in\mathcal{N}_i}\pi^{(ij)}\textbf{P}_{n,\ell}^j
    \end{aligned}\rightarrow \pi^{(ij)}\geq 0, \sum_{j\in\mathcal{N}_i}\pi^{(ij)} = 1 \label{Eq31}
\end{align}
and the coupled $(\hat{\textbf{x}}_{n}^i,\textbf{P}_{n}^i)_{i\in\mathcal{N}}$ is then achieved if the formula $\Pi= \pi^{(1,j)} \mathcal{R}^n$ is primitive
\begin{align*}
    \hat{\textbf{x}}_{n,\ell+1} &= \left(\Pi\otimes I\right)\hat{\textbf{x}}_{n,\ell}\\
    &= \left(\Pi\otimes I\right)\dots\left(\Pi\otimes I\right)\hat{\textbf{x}}_{n,0} = \left(\Pi^{\ell+1}\otimes I\right)\hat{\textbf{x}}_{n,0}
\end{align*}
and therefore,
\begin{align*}
    \lim_{\ell\to\infty}\left(\Pi^{\ell+1}\right) = 1\textbf{v}^\top
\end{align*}
where as $\ell\to\infty$ with the column vector of $\textbf{v}$,
\begin{align*}
    \hat{\textbf{x}}_{n,\ell+1} = \left(1\textbf{v}^\top\otimes I\right)\hat{\textbf{x}}_{n,0}
\end{align*}
the estimated state and the error covariance matrix would be,
\begin{gather*}
    \hat{\textbf{x}}_{n,\ell+1} = v_1\hat{\textbf{x}}_{n,0}^1 + v_1\hat{\textbf{x}}_{n,0}^2 + \dots + v_1\hat{\textbf{x}}_{n,0}^k = \hat{\textbf{x}}_{n}^\ast\\
    \textbf{P}_{n,\ell+1} = v_1\textbf{P}_{n,0}^1 + v_1\textbf{P}_{n,0}^2 + \dots + v_1\textbf{P}_{n,0}^k = \textbf{P}_{n}^\ast
\end{gather*}
Furthermore, the algorithm of the weighted average consensus from Eq.(\ref{EqB1}) o Eq.(\ref{EqB5}) is written below as the extension of the standard UKF, 
\begin{enumerate}
    \item For every $i\in\mathcal{N}$, collect the information of $\hat{\textbf{y}}_n^{(i)}$ and find
    \begin{align}
        \hat{\textbf{x}}_n^i &= \hat{\textbf{x}}_n + \textbf{W}_n z_n \longrightarrow z_n = \textbf{y}_n - \hat{\textbf{y}}_n^{(i)}\tag{$\beta_1$}\label{EqB1}\\
        \textbf{P}_n^i &= \hat{\textbf{P}}_{n} - \textbf{W}_n\hat{\textbf{S}}_{n}\textbf{W}_n^\top \tag{$\beta_2$}
    \end{align}
    \item Initialize that $\hat{\textbf{x}}_n^i = \hat{\textbf{x}}_{n,0}^i$ and $\textbf{P}_{n,0}^i = \textbf{P}_{n,0}^i$
    \item For the $\ell = 0,1,\dots,l$, apply the method of the weighted average consensus, such that:
    \begin{enumerate}
        \item Broadcast the coupled node data $\hat{\textbf{x}}_{n,\ell}^i$ and $\textbf{P}_{n,\ell}^i$ to the surrounding neighborhoods $j\in\mathcal{N}_i \setminus (i)$
        \item Ensure the information of $\hat{\textbf{x}}_{n,\ell}^j$ and $\textbf{P}_{n,\ell}^j$ from the whole neighborhoods $j\in\mathcal{N}_i \setminus (i)$
        \item Collect the coupled data $\hat{\textbf{x}}_{n,\ell}^j$ and $\textbf{P}_{n,\ell}^j$ based on
    \end{enumerate}
    \begin{align}
    \left(\hat{\textbf{x}}_{n,\ell+1}^i, \textbf{P}_{n,\ell+1}^i\right) = \sum_{j\in\mathcal{N}_i}\pi^{(ij)}\left(\hat{\textbf{x}}_{n,\ell}^j, \textbf{P}_{n,\ell}^j\right) 
    \tag{$\beta_3$}
    \end{align}
    \item Setup the estimated state as, 
    \begin{align}
        \hat{\textbf{x}}_n^i = \hat{\textbf{x}}_{n+l}^i \quad\textrm{and}\quad \textbf{P}_n^i = \textbf{P}_{n+l}^i \tag{$\beta_4$}
    \end{align}
    \item Perform the updated of the prediction error, therefore
    \begin{align}\begin{aligned}
        \hat{\textbf{x}}_{a,n}^+ &= \sum_{i=0}^{2L} W^{(i)} \hat{\chi}_{a,n}^{(i)}\\
        \hat{\textbf{P}}_{a,n}^+ &= \sum_{i=0}^{2L} W^{(i)}\left[\hat{\chi}_{a,n}^{(i)} - \hat{\textbf{x}}_{a,n}\right]\left[\hat{\chi}_{a,n}^{(i)} - \hat{\textbf{x}}_{a,n}\right]^\top
    \end{aligned}\tag{$\beta_5$}\label{EqB5}\end{align}
\end{enumerate}
As for theoretical feasibility of the stochastic boundedness with in-depth mathematical proof is well-explained in \cite{R30}

\section{Numerical Designs and Findings}\label{Sec6}
\begin{figure}[h!]
    \centering
    \includegraphics[width = .9\linewidth]{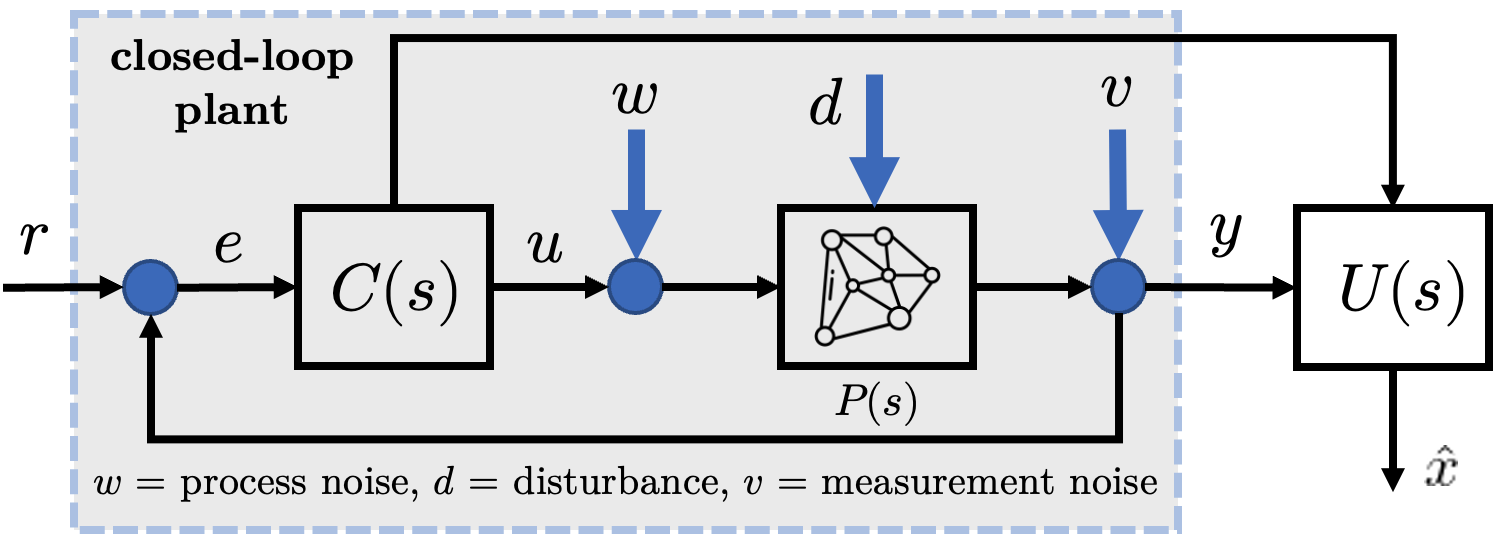}
    \caption{Block diagram of the system}
    \label{F9}
\end{figure}

\begin{figure*}[t!]
     \centering
     \begin{subfigure}[b]{0.325\textwidth}
         \centering
         \includegraphics[width=\textwidth]{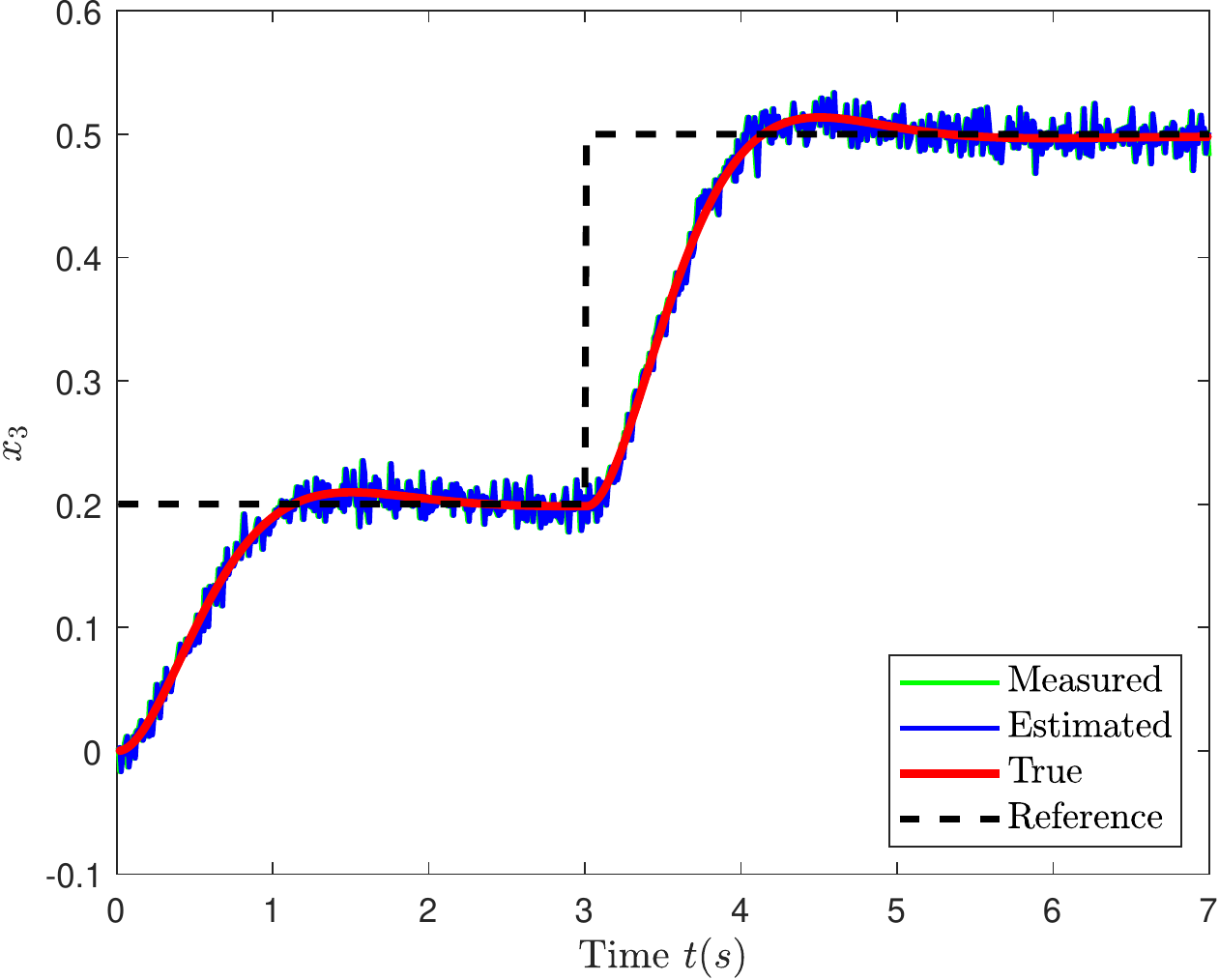}
         \caption{States comparison of $x_3$}
         \label{P3a}
     \end{subfigure}\qquad
     \begin{subfigure}[b]{0.33\textwidth}
         \centering
         \includegraphics[width=\textwidth]{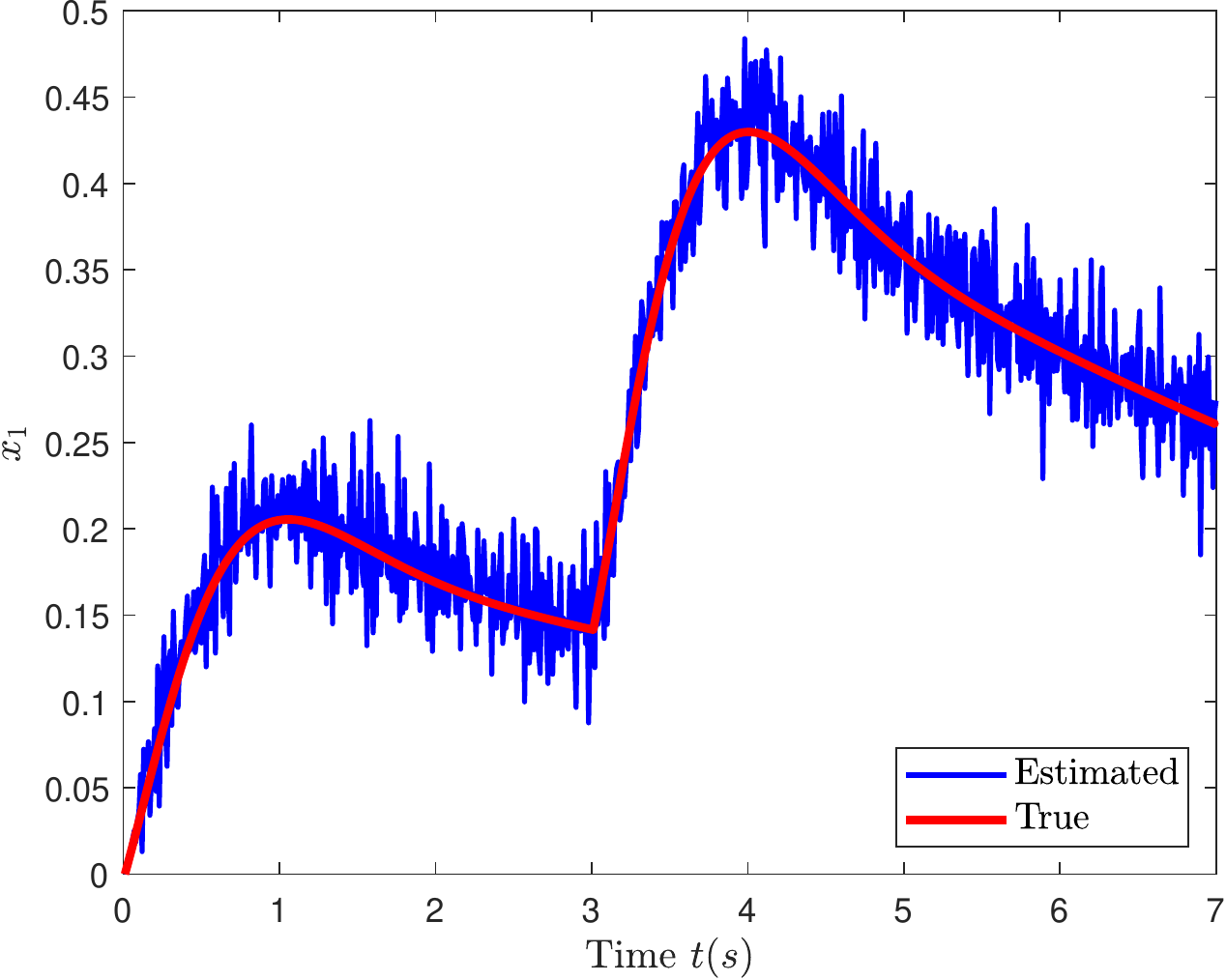}
         \caption{States comparison of $x_1$}
         \label{P3b}
     \end{subfigure}
     \caption{The performance results of pitch angle ($x_3 = \theta$) using plant $\Phi_3$ of the measured ($y$), the proposed estimated ($x_u$) and the true ($x_t$) states according to the reference ($r$) along with the free-state}
     \label{P3}
\end{figure*}

Having discussed the various dynamical systems denoted as $P(s)$ along with some control scenarios $C(s)$ and their stability, the network is portrayed as in Fig.(\ref{F9}) considering the process $w$ and measurement $v$ noise and the disturbance $d$. These control scenarios are diverged from the classical perspective (PID, LQR, etc) to the modern actor-critic reinforcement learning and other dynamic control to match the condition of the system.
Furthermore, the modified UKF estimation $U(s)$ is then applied to touch the hidden states of the systems to reach the information leading to sensorless design. This estimation is then compared to other centralized and distributed filtering as explained in the followings. The whole dynamics of the tested systems are constructed according to certain plants $\Phi_n(s), \forall n = 1\to 4$ as shown in Fig.(\ref{C}). As for the first $\Phi_1$, the mass $m$ and the damping parameter $b$ are set to be 1000 kg and 50 N.s/m in turn while the nominal control $u$ is 500 N with dynamic reference $v_r$ in 10 m/s and 7 m/s. The second plant $\Phi_2$ comes up with a quarter-body mass $M_1$ as 2500 kg and the suspension mass $M_2$ as 320 kg whereas the parameters of the spring with respect to system $k_1$ and wheel $k_2$ are 80,000 N/m and 500,000 N/m in turn while the damping of the same respected terms, $b_1$ and $b_2$, have 350 N.s/m and 15,020 N.s/m respectively. Regarding the third plant $\Phi_3$, according to Eq.(\ref{Eq15}), the values of $c_k, \forall k = 1\to 7$ constitute $-0.313$, $56.7$, $0.232$, $-0.0139$, $-0.426$, $0.0203$, and $56.7$ while the fourth $\Phi_4$ with speed $\dot{\theta}$ and position $\theta$ comprises the variables as follows. The moment $J$ is 0.01 kg.m$^2$ with the friction parameter $b$ of 0.1 N.m.s and the same gain $\kappa$ makes of 0.01 to the system resistance $R$ equals to 1 Ohm and inductance $L$ with 0.5 H. Regarding the parameters of the proposed estimation methods, the time-sampling ($t_s$), covariance matrices of $Q$ and $R$ are
\begin{figure*}[t!]
     \centering
     \begin{subfigure}[b]{0.325\textwidth}
         \centering
         \includegraphics[width=\textwidth]{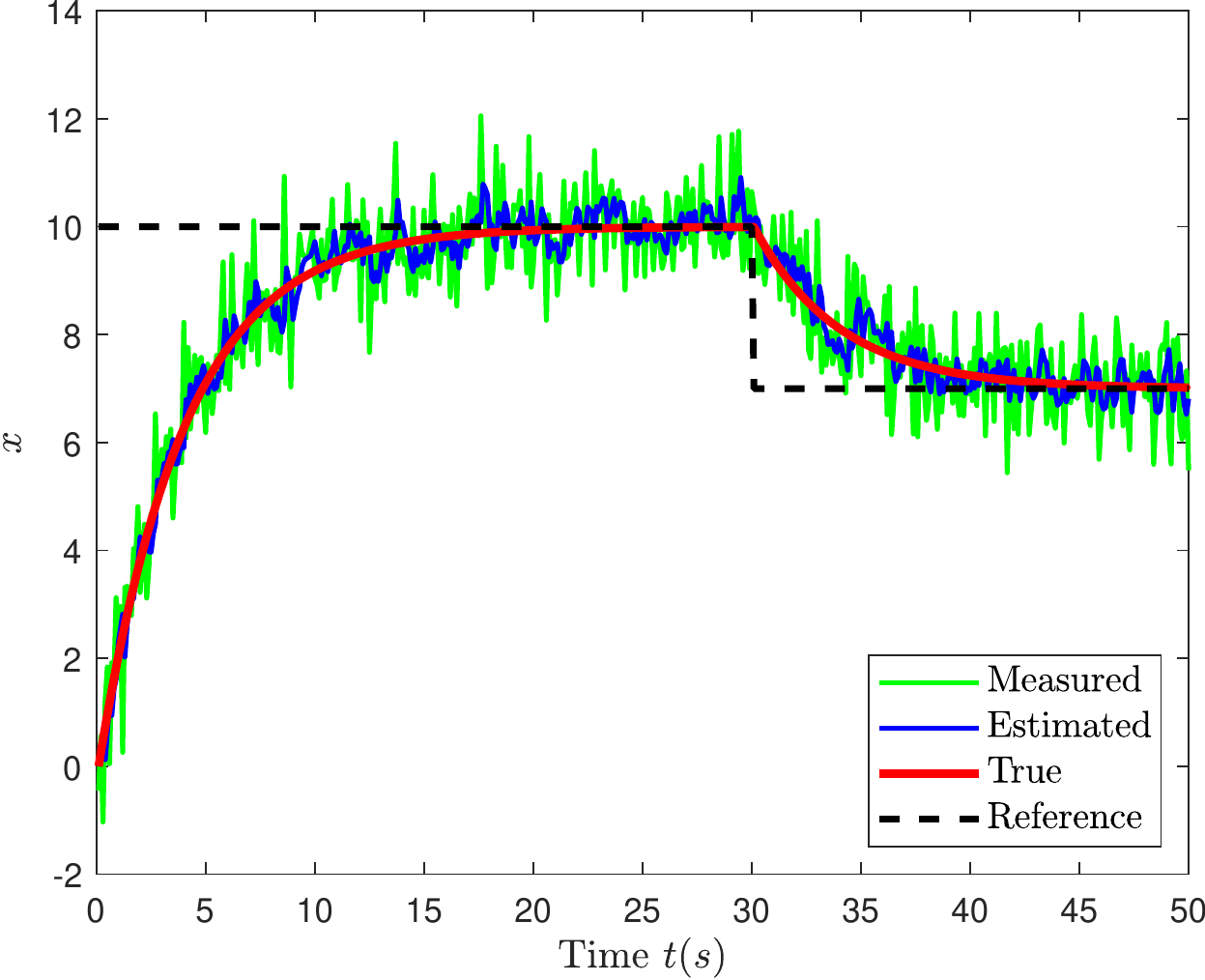}
         \caption{States comparison over certain reference points}
         \label{P1a}
     \end{subfigure}\qquad
     \begin{subfigure}[b]{0.33\textwidth}
         \centering
         \includegraphics[width=\textwidth]{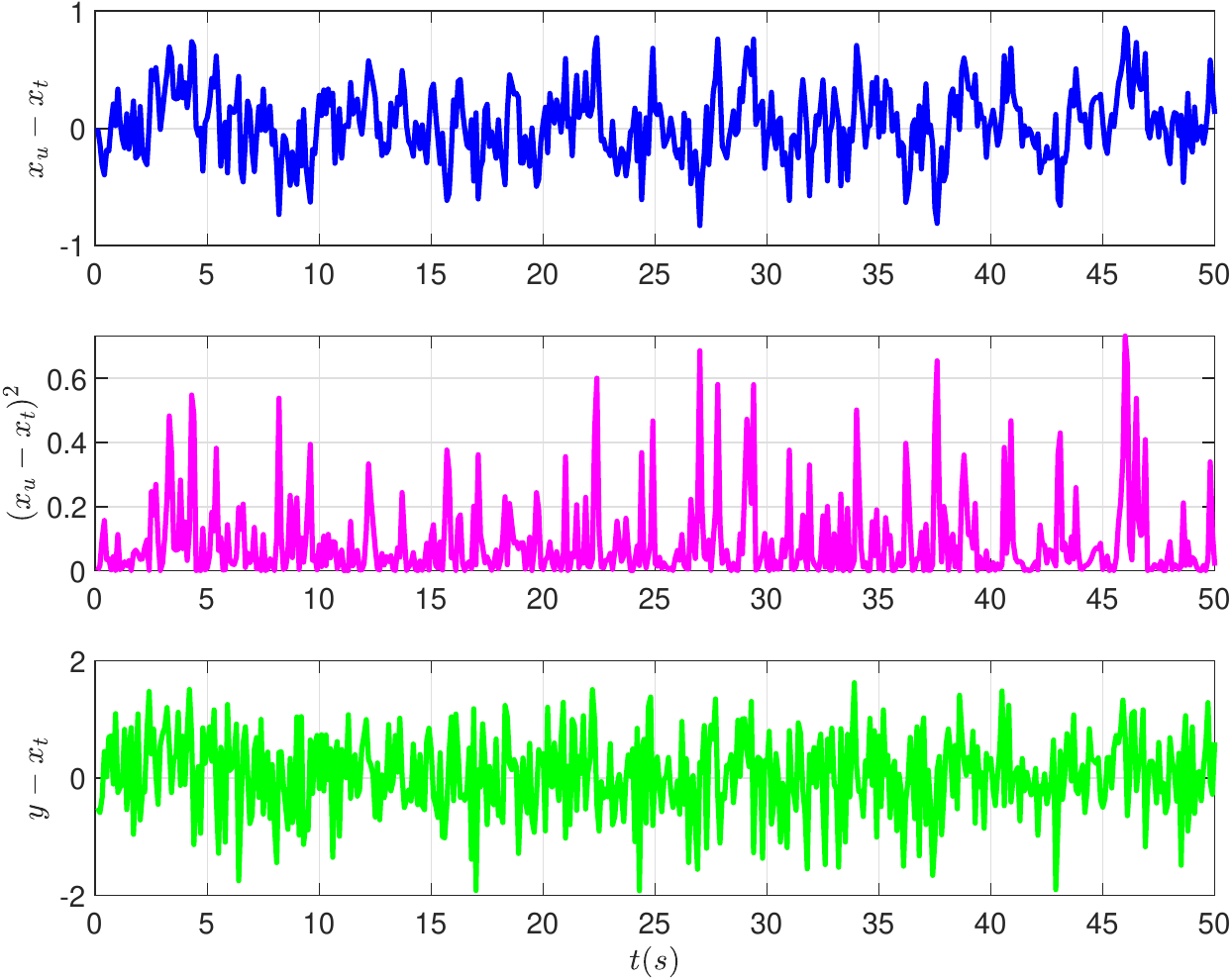}
         \caption{Error comparisons of the states}
         \label{P1b}
     \end{subfigure}
     \caption{(a) The performance results of speed ($x = v$) using plant $\Phi_1$ of the measured ($y$), the proposed estimated ($x_u$) and the true ($x_t$) states according to the reference ($r$); (b) while it shows the error comparisons of the measured and the estimated over the true values}
     \label{P1}
\end{figure*}
\begin{figure*}[t!]
     \centering
     \begin{subfigure}[b]{0.33\textwidth}
         \centering
         \includegraphics[width=\textwidth]{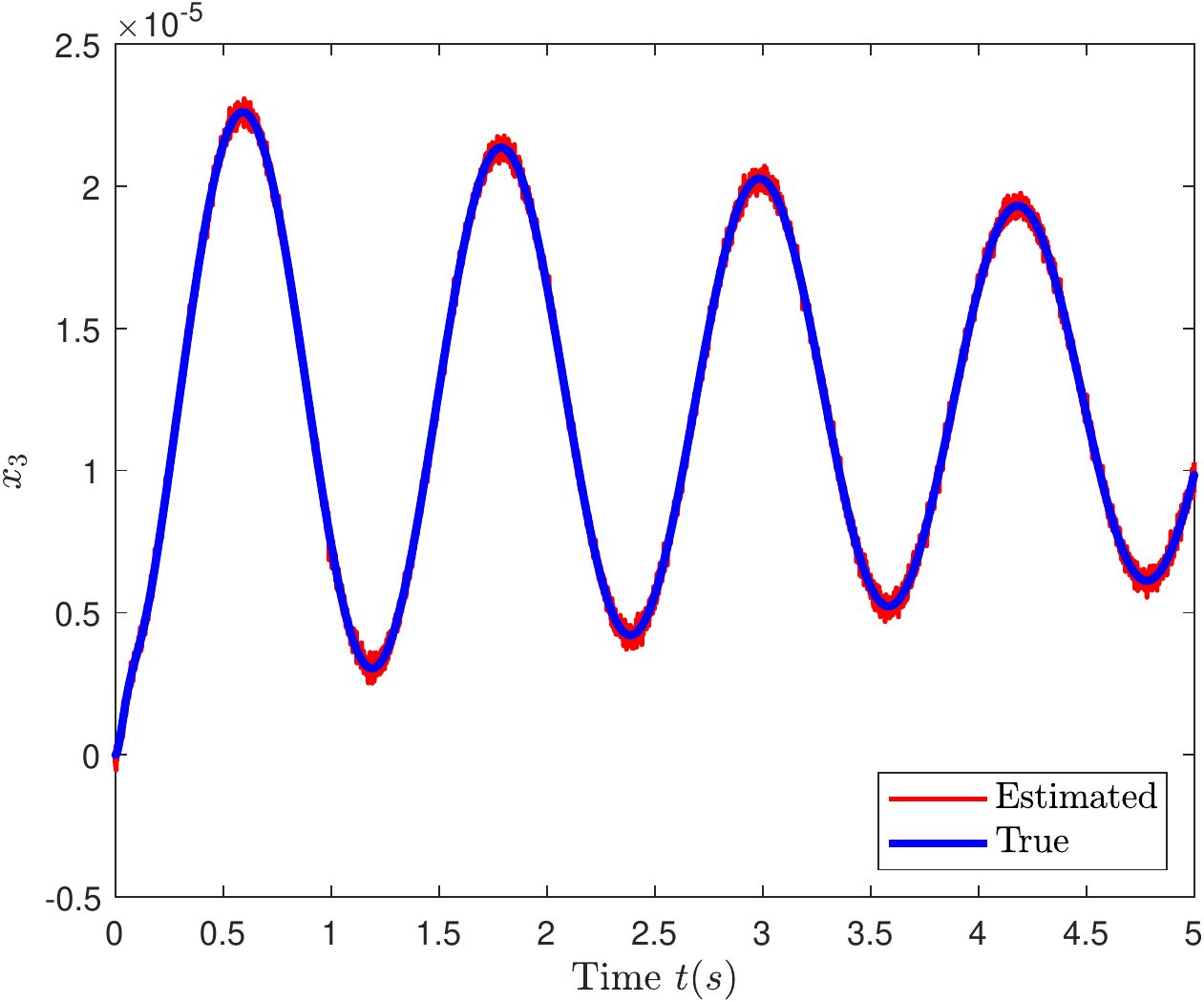}
         \caption{State ($x_3$) response due to disturbance}
         \label{P2a}
     \end{subfigure}\qquad
     \begin{subfigure}[b]{0.325\textwidth}
         \centering
         \includegraphics[width=\textwidth]{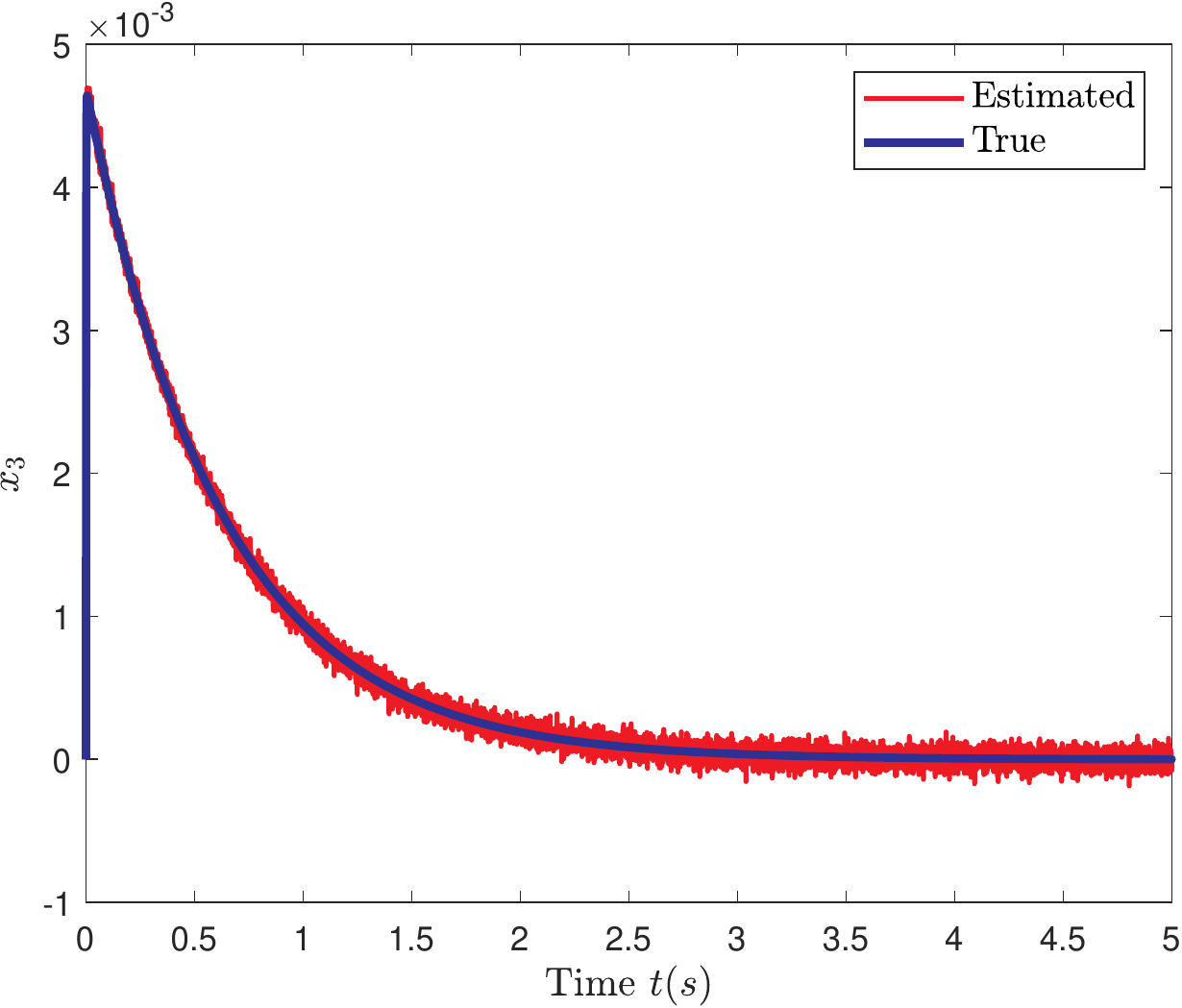}
         \caption{State ($x_3$) response due to disturbance}
         \label{P2b}
     \end{subfigure}\\[1em]
     \begin{subfigure}[b]{0.33\textwidth}
         \centering
         \includegraphics[width=\textwidth]{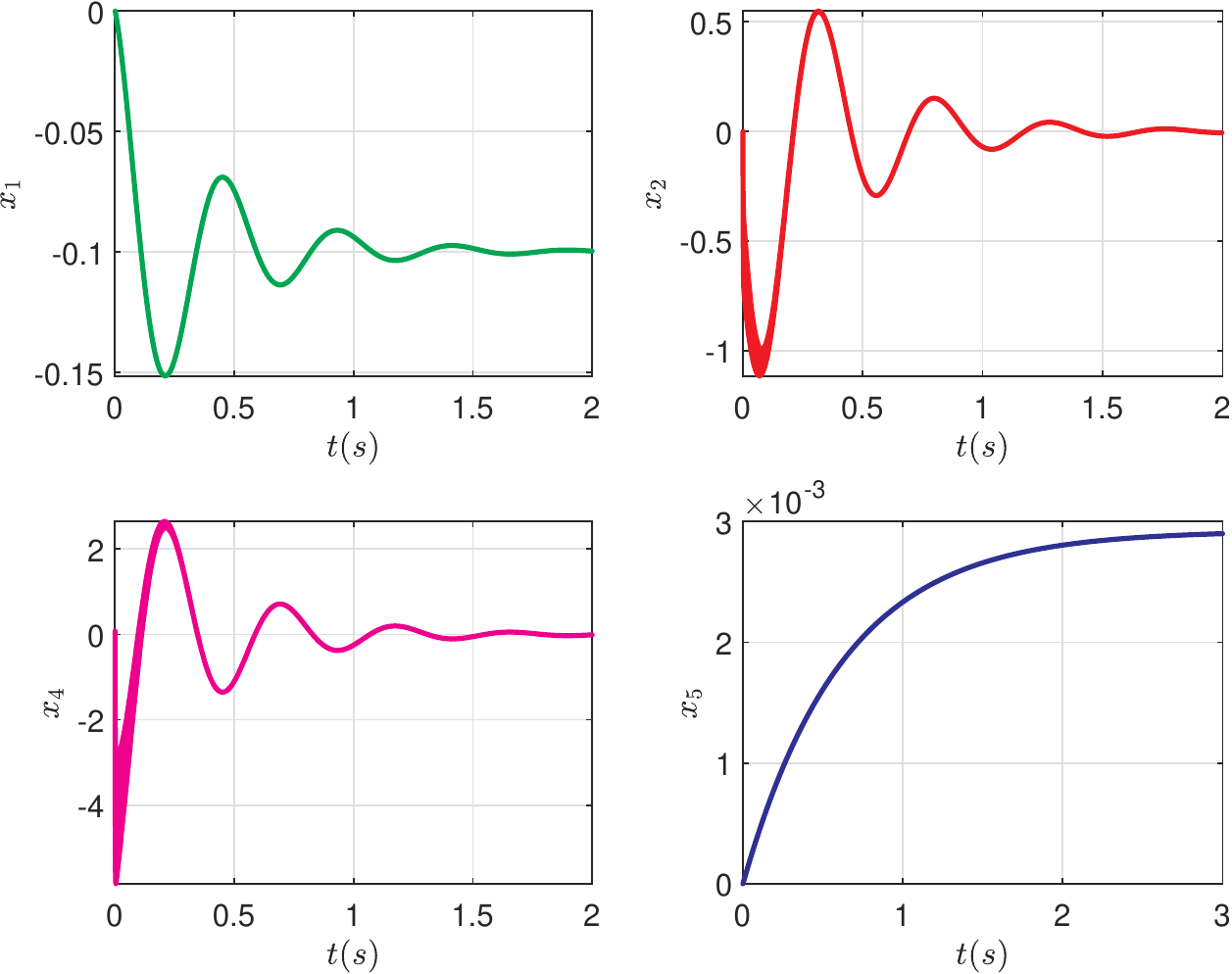}
         \caption{The closed-loop responses of the states}
         \label{P2c}
     \end{subfigure}\quad
     \begin{subfigure}[b]{0.35\textwidth}
         \centering
         \includegraphics[width=\textwidth]{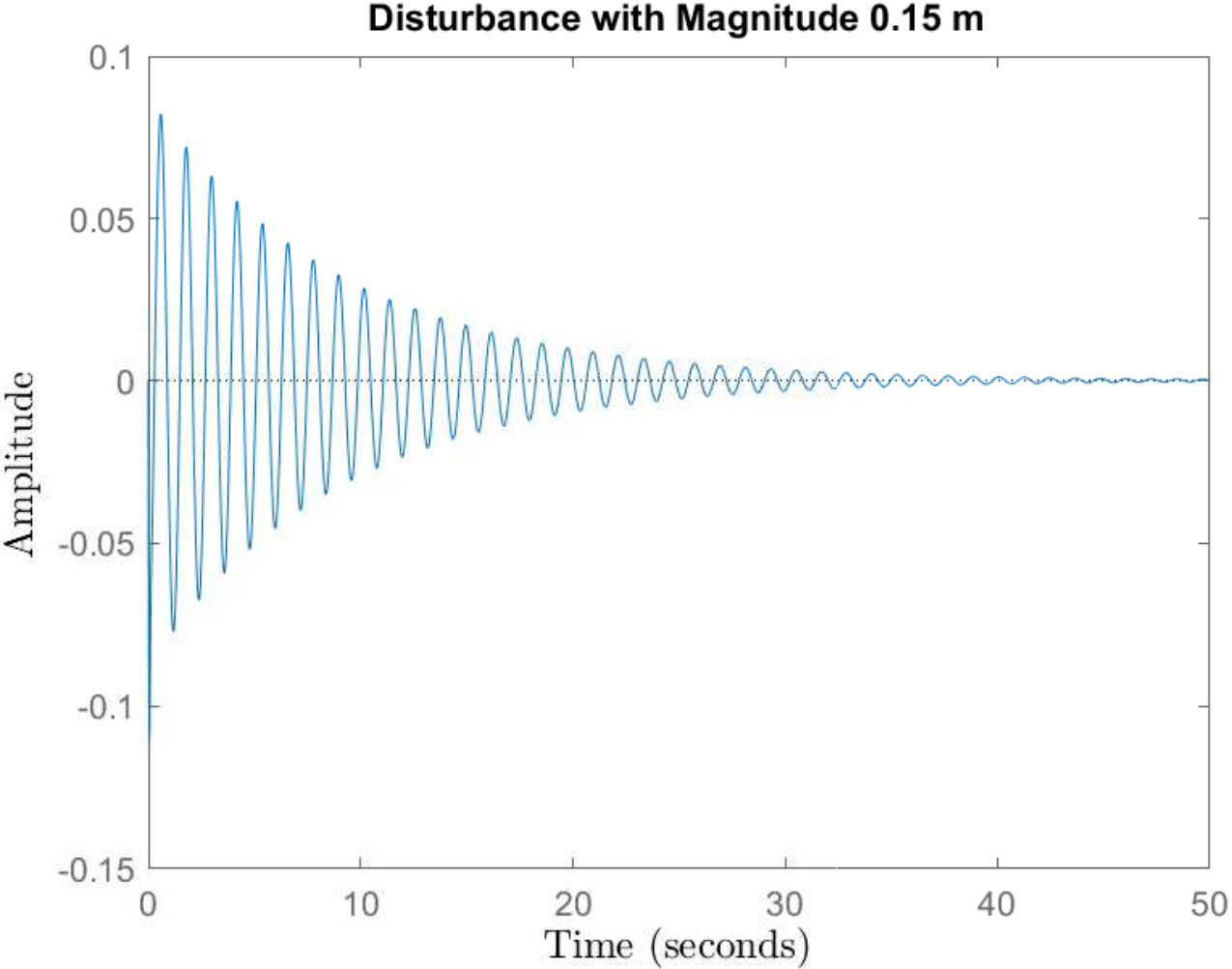}
         \caption{The open-loop response with 0.15 m disturbance}
         \label{P2d}
     \end{subfigure}
     \caption{(a), (b) The performance results of ($x_3 = \varphi$) using plant $\Phi_2$ of the proposed estimated ($x_u$) and the true ($x_t$) with some disturbance ($\gamma$); (c) while it highlights the closed-loop response of the states ($x_1, x_2, x_4, x_5$);  and (d) the open-loop response if $\gamma$ is applied to the system} 
     \label{P2}
\end{figure*}
\begin{align*}
    \begin{array}{l|l|l|l}
    \textrm{Plant} & t_s & R & Q \\[.2em] \hline
    \Phi_1 = & 0.01 & 0.5 & 0.1\\[.2em]
    \Phi_2 = & 0.0005 & 0.05 & 10\\[.2em]
    \Phi_3 = & 0.01 & 1 & pH^\top H\\[.2em]
    \Phi_4 = & 0.01 & 1 & pH^\top H
    \end{array} 
\end{align*}
where the measurement noise $v$ of the every system $\Phi_n(s)$ is formulated based on the output matrix $H$ with some unique distribution $\mathcal{R}_u\in(0,1)$ by
\begin{align*}
    v = \sqrt{R}\times \textrm{randn}[\textrm{size}(H,1),1]
\end{align*}
where (randn) means the normal distribution of GRV while the $\textrm{size}(\epsilon_1, \epsilon_2)$ denotes either the row or column dimension $\epsilon_2\in(1,2)$ in turn of the matrix $\epsilon_1$. $p_0$ as the arbitrary constant solving the algebraic Riccati equation (ARE) equals to 50 with eye($\bullet$) and discrete($\bullet$) states the identity matrix of length ($\bullet$), known as the maximum size or dimension, and the discretized systems of ($\bullet$), such that it shows
\begin{align*}
    \textbf{P}_0 = p_0\times \textrm{eye}(F)\longrightarrow F = \textrm{discrete}(A)
\end{align*}
while the performance of the estimation error is approached with the error as opposed to the true values, having no noise in the systems, 
\begin{align*}
    e = x_u - x_t
\end{align*}
Regarding the performance of plant $\Phi_1$, the desired references in 50 s are situated in two different speed values and the control along with the estimated values could deal with the changes as depicted in Fig.(\ref{P1a}) while the performance errors are written in Fig.(\ref{P1b}) showing the zero convergence. As for the second plant $\Phi_2$, the dynamics of the disturbance-effect systems, the closed-loop and the open-loop are discussed, saying the capability of handling the external forces in under 5 seconds. The values $x_3\to y_1$ between the true and the estimation parallels with slight difference on the peak and trough as illustrated in Fig.(\ref{P2a}) while the closed-loop in Fig.(\ref{P2b}) also performs almost on par with preceding dynamics. Fig.(\ref{P2c}) and Fig.(\ref{P2d}) constitute the stabilized and the non-control performance of the systems. Furthermore, the third dynamics $\Phi_3$ describes the control of the pitch angle $\theta$ of an aircraft and it is built into two divergent reference points, 0.2 and 0.5, within 7 seconds. In terms of control scenario, the systems could trace the reference while for the estimated states, the controlled state $x_3$ in Fig.(\ref{P3b}) and the free-state $x_1$ in Fig.(\ref{P3b}) are well-covered by the proposed algorithm as opposed to the true states.
\begin{figure*}[t!]
     \centering
     \begin{subfigure}[b]{0.33\textwidth}
         \centering
         \includegraphics[width=\textwidth]{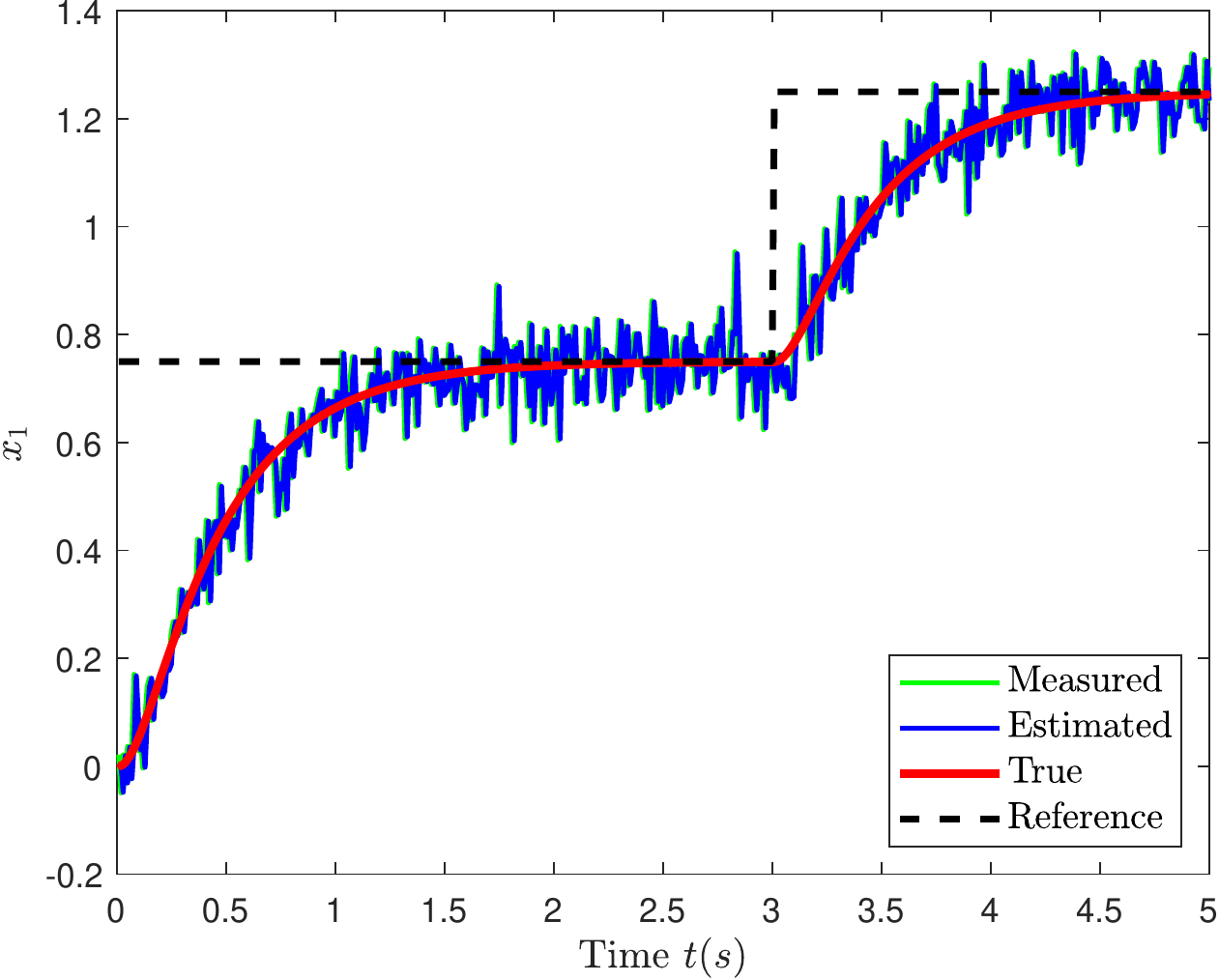}
         \caption{}
         \label{P4a}
     \end{subfigure}\qquad
     \begin{subfigure}[b]{0.33\textwidth}
         \centering
         \includegraphics[width=\textwidth]{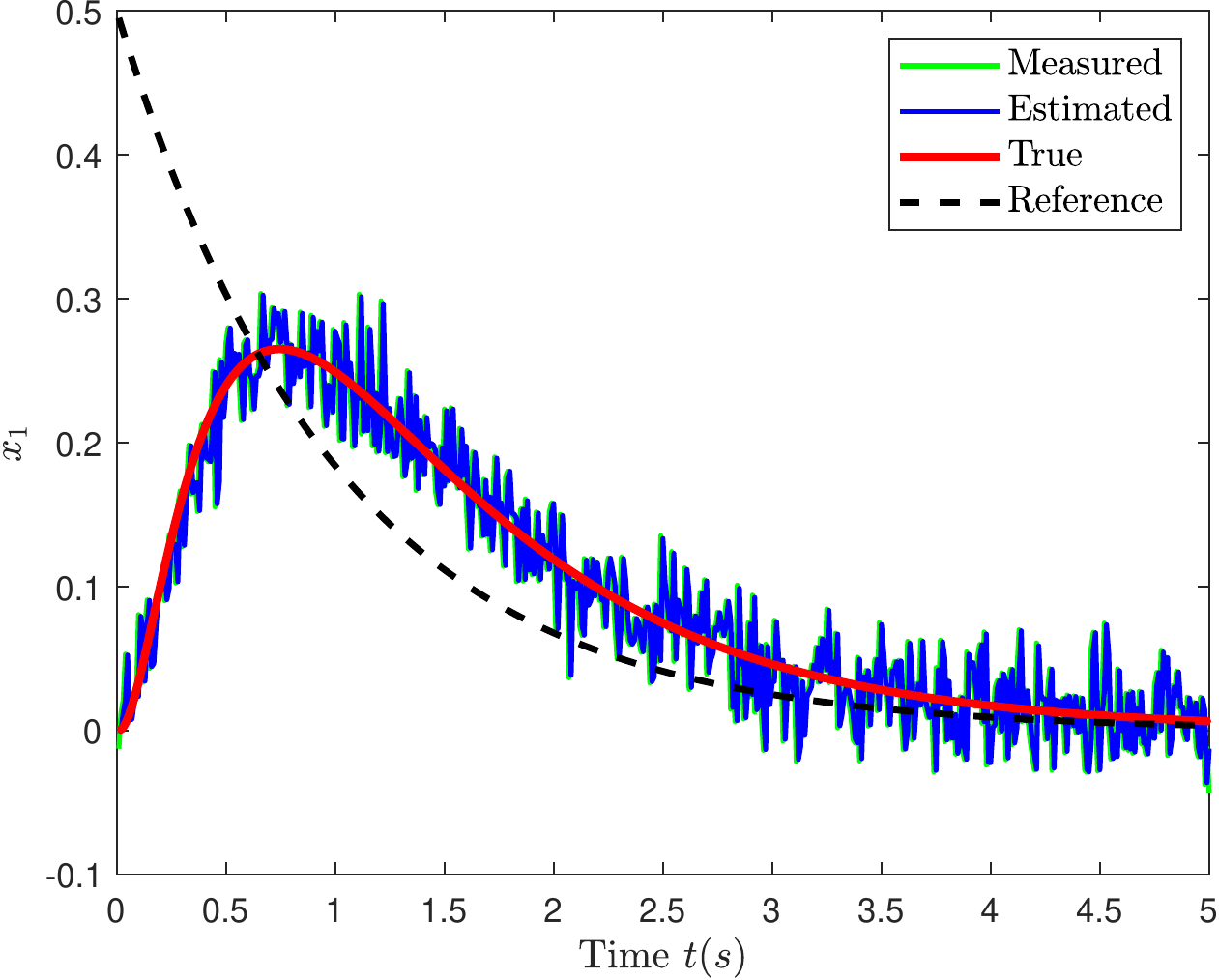}
         \caption{}
         \label{P4b}
     \end{subfigure}
     \caption{The performance results of speed ($x_1 = \dot{\theta}$) in various dynamics using plant $\Phi_4$ of the measured ($y$), the proposed estimated ($x_u$) and the true ($x_t$) states according to the reference ($r$)}
     \label{P4}
\end{figure*}
\begin{figure*}[t!]
     \begin{subfigure}[b]{0.33\textwidth}
         \centering
         \includegraphics[width=\textwidth]{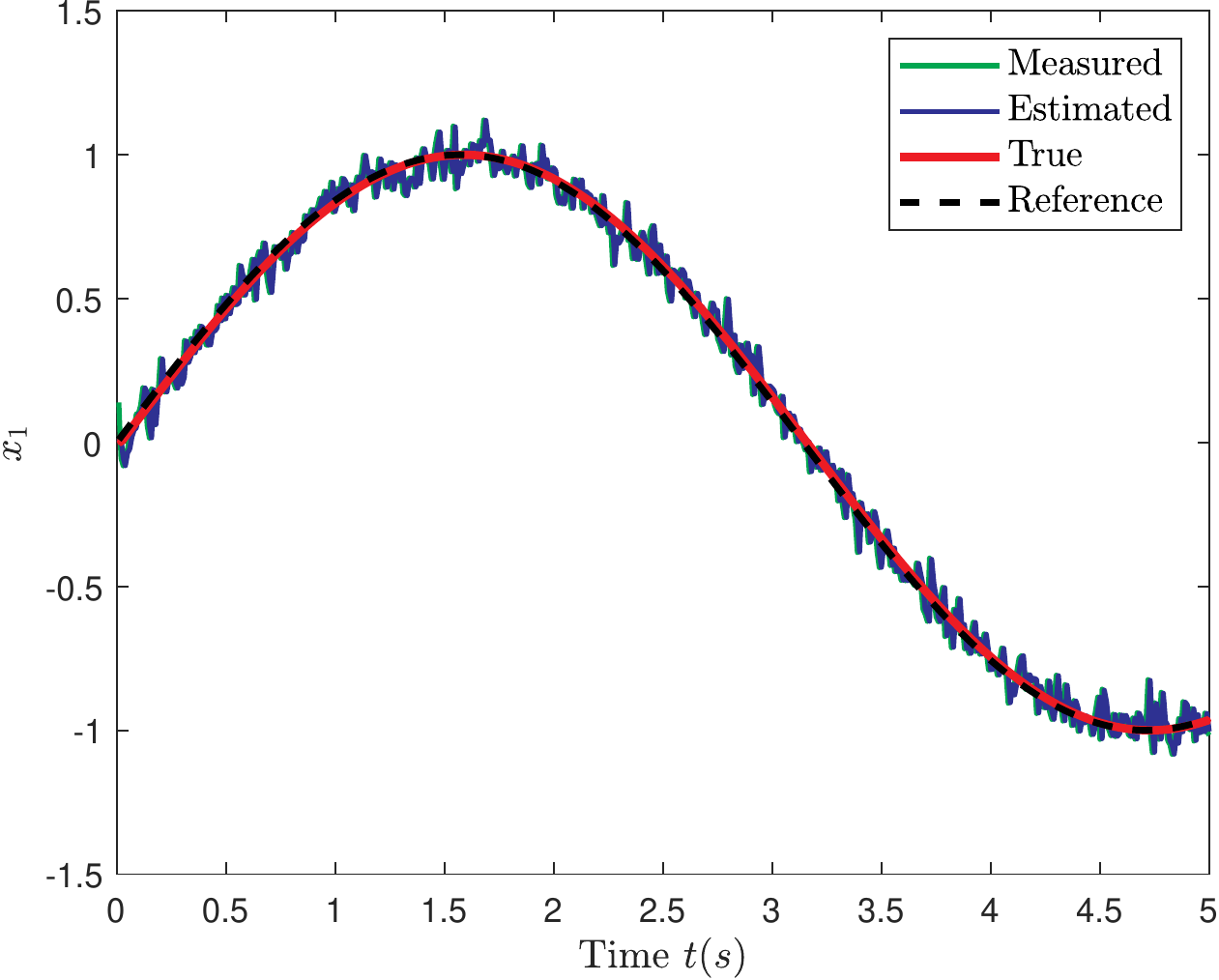}
         \caption{}
         \label{P5a}
     \end{subfigure}
     \begin{subfigure}[b]{0.33\textwidth}
         \centering
         \includegraphics[width=\textwidth]{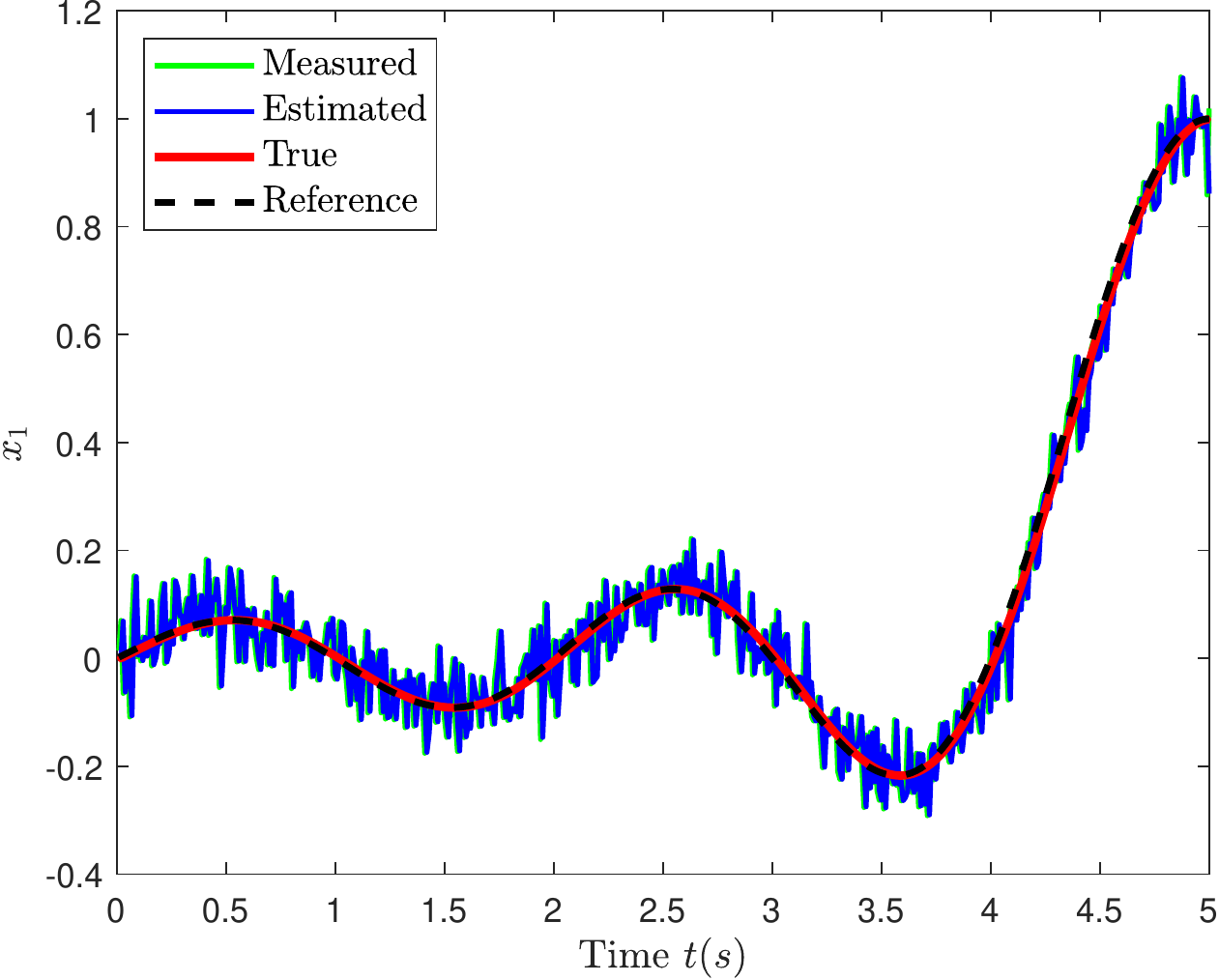}
         \caption{}
         \label{P5b}
     \end{subfigure}
     \begin{subfigure}[b]{0.33\textwidth}
         \centering
         \includegraphics[width=\textwidth]{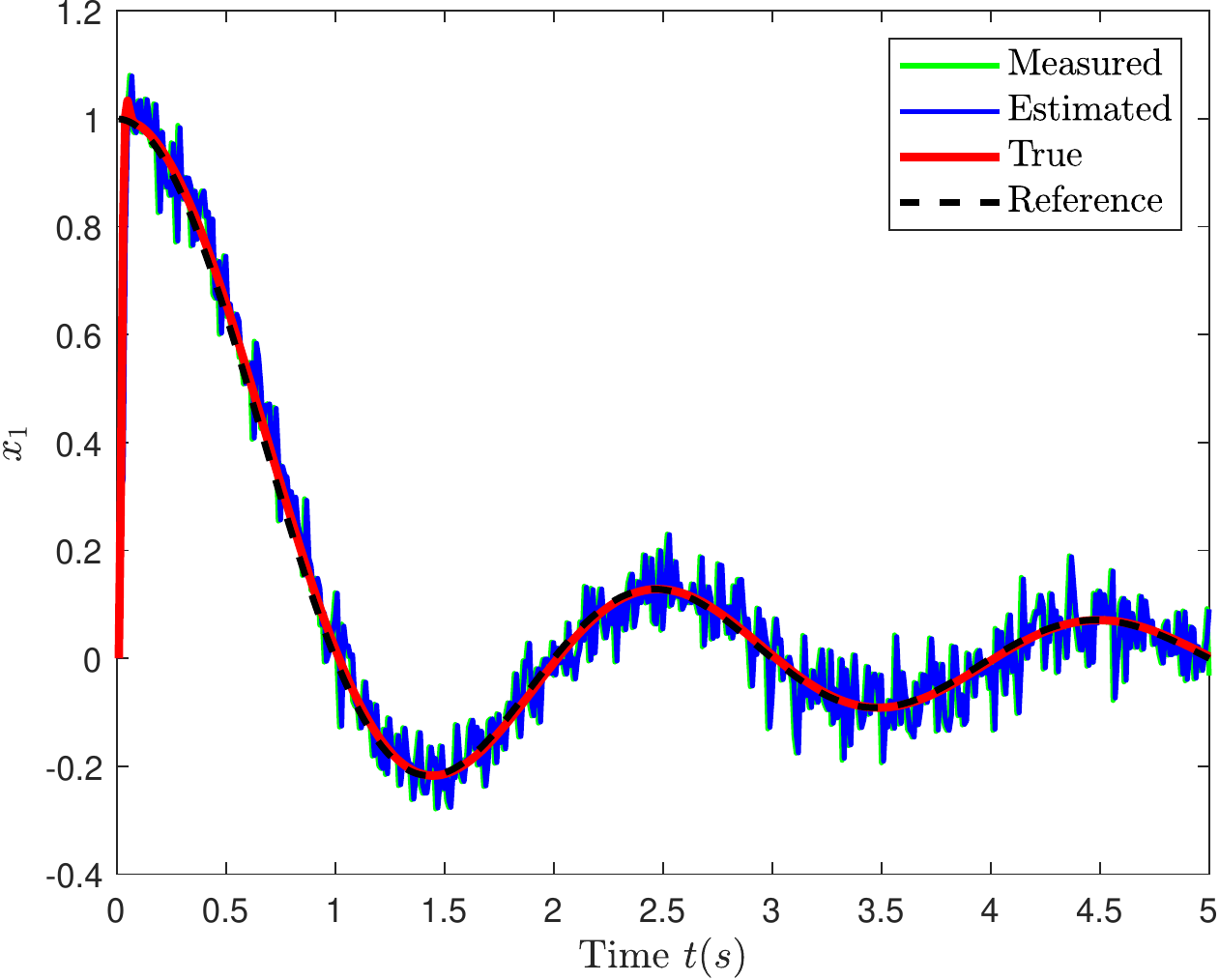}
         \caption{}
         \label{P5c}
     \end{subfigure}
     \caption{The performance results of speed ($x_1 = \theta$) in various dynamics using plant $\Phi_4$ of the measured ($y$), the proposed estimated ($x_u$) and the true ($x_t$) states according to the reference ($r$)}
     \label{P5}
\end{figure*}
Likewise, the rotor dynamics $\Phi_4$ for speed variable $\dot{\theta}$ with two different reference scenarios in Fig.(\ref{P4a}) and Fig.(\ref{P4b}) are handled by the control design and well-estimated by the weighted average consensus method. Finally, the position-type variable $\theta$ of $\Phi_4$ comprises the same trends with respect to the comparison of the measured, the estimated and the true states with various dynamics as portrayed in Fig.(\ref{P5}). To conclude, the control designs from various plants $\Phi_n, \forall n = 1\to 4$ successfully capture the dynamics of the systems while the proposed algorithm effectively tracks the true values under some disturbances this also leads to the sensorless design of the future works. The performances are affected solely on the covariance matrix pairs $(Q,R)$ and the measurements. Beyond that, the performances of the estimation are also weighed according to the reviews of distributed estimation \cite{R48,R49,R57} along with various applications in terms of non-linear uncertainties interconnected system \cite{R50,R51,R52,R53,R54,R55}, and Pareto optimization \cite{R56}. The presented results from arbitrary initial conditions ($x_0$), in terms of estimation errors, highlight the convergence, almost identical as the centralized Kalman with different in the noise scalability, as compared to distributed estimation studied in \cite{R23,R49,R57}. For the more severe faulty unobserved systems, the mechanism to maturely detect the states is highly required while for large-scale systems, the information of purely local measurement is also possible as opposed to the local and its neighborhood.

\section{Conclusions}\label{Sec7}
The mathematical dynamics of the vehicle systems along with the graphs have been constructively designed under some disturbance as the object of the performance results. The control scenarios for the tested plants $\Phi_n$ along with some stability analysis have also been discussed comprehensively to check the observability of the systems. The standard UKF and the proposed of the weighted average consensus estimation method is written considering some neighborhoods events as the local collecting information to obtain the more accurate estimated states. The results of the designs conclude the effectiveness of the control designs along with the proposed estimation algorithm to track the hidden states. For further research, the sensorless design to vehicles dynamics is elaborated leading to the autonomous concepts along with some fault-tolerant learning control if faults occur to negate the failured systems.

\section*{Acknowledgment}
This research was provided by a funding granted by the Engineering Physics Department of Institut Teknologi Sepuluh Nopember (ITS), Indonesia with letter contract number: 1868/PKS/ITS/2022 in May 24, 2022. We thank our colleagues for the ideas, dedication, and times for the final paper
\newpage
\bibliographystyle{ieeetr}
\bibliography{EN-Bibliography}
\newpage
\end{document}